\documentclass[journal]{IEEEtran}
\usepackage{hyperref}

\usepackage{cite}

\ifCLASSINFOpdf
  \usepackage[pdftex]{graphicx}
  \graphicspath{{./figures/}}
\else
  \usepackage[dvips]{graphicx}
\fi
\usepackage{amsmath}
\usepackage{amsfonts}
\ifCLASSOPTIONcompsoc
  \usepackage[caption=false,font=normalsize,labelfont=sf,textfont=sf]{subfig}
\else
  \usepackage[caption=false,font=footnotesize]{subfig}
\fi

\usepackage[utf8]{inputenc}
\usepackage[capitalise]{cleveref}

\usepackage{multirow}

\usepackage[nomargin,inline,final]{fixme}
\fxusetheme{color}

\newsavebox{\measurebox}

\usepackage[utf8]{inputenc}
\usepackage{pgfplots}
\DeclareUnicodeCharacter{2212}{−}
\usepgfplotslibrary{groupplots,dateplot}
\usetikzlibrary{patterns,shapes.arrows}
\pgfplotsset{compat=newest}
\newlength\figureheight
\newlength\figurewidth

\title{OSLO: On-the-Sphere Learning for Omnidirectional images and its application to 360-degree image compression}

\begin{document}

\author{Navid Mahmoudian Bidgoli,
        Roberto G. de A. Azevedo,
        Thomas Maugey, 
        Aline Roumy, 
        Pascal Frossard%
\thanks{This work was supported by the Cominlabs excellence laboratory with funding from the French National Research Agency (ANR-10-LABX-07-01), InterCom and MOVE projects.}
\thanks{Navid Mahmoudian Bidgoli, Thomas Maugey, Aline Roumy are with Inria, Univ Rennes, CNRS, IRISA, France (e-mails: \{navid.mahmoudian-bidgoli, thomas.maugey, aline.roumy\}@inria.fr).}
\thanks{Roberto G. de A. Azevedo is with ETH Zurich.}
\thanks{Pascal Frossard is with the Signal Processing Laboratory (LTS4), École Polytechnique Fédérale de Lausanne (EPFL).}
\thanks{This work was developed while Roberto G. de A. Azevedo was with the Signal Processing Laboratory (LTS4), École Polytechnique Fédérale de Lausanne (EPFL).}}


\maketitle

\begin{abstract}

State-of-the-art 2D image compression schemes rely on the power of convolutional neural networks (CNNs). Although CNNs offer promising perspectives for 2D image compression, extending such models to omnidirectional images is not straightforward. First, omnidirectional images have specific spatial and statistical properties that can not be fully captured by current CNN models. Second, basic mathematical operations composing a CNN architecture, \emph{e.g.}, translation and sampling, are not well-defined on the sphere. In this paper, we study the learning of representation models for omnidirectional images and propose to use the properties of HEALPix uniform sampling of the sphere to redefine the mathematical tools used in deep learning models for omnidirectional images. In particular, we: i)~propose the definition of a new convolution operation on the sphere that keeps the high expressiveness and the low complexity of a classical 2D convolution; ii)~adapt standard CNN techniques such as stride, iterative aggregation, and pixel shuffling to the spherical domain; and then iii)~apply our new framework to the task of omnidirectional image compression. Our experiments show that our proposed on-the-sphere solution leads to a better compression gain that can save 13.7\% of the bit rate compared to similar learned models applied to equirectangular images. Also, compared to learning models based on graph convolutional networks, our solution supports more expressive filters that can preserve high frequencies and provide a better perceptual quality of the compressed images. Such results demonstrate the efficiency of the proposed framework, which opens new research venues for other omnidirectional vision tasks to be effectively implemented on the sphere manifold.
\end{abstract}


%
\IEEEpeerreviewmaketitle

\section{Introduction}
360-degree, or \emph{omnidirectional}, visual contents are high-resolution \textit{spherical} images/videos that capture a 3D scene with a full field of view. By allowing users to adjust their viewing orientation freely, omnidirectional contents support an immersive and interactive experience. Due to their unique features, omnidirectional images have become increasingly popular in different applications such as Virtual Reality~(VR)~\cite{hou2021predictive,nam2021an}, augmented reality~(AR), and autonomous driving~\cite{deng2020restricted}, which at the same time outlines the crucial need for effective processing tools. Many recent imaging techniques are based on Convolutional Neural Networks~(CNNs), which have achieved great success for planar images in various computer vision and image processing applications such as image classification or compression. However, their extension to omnidirectional images raises several issues. 

\begin{figure}
\centering
\includegraphics[width=0.8\linewidth]{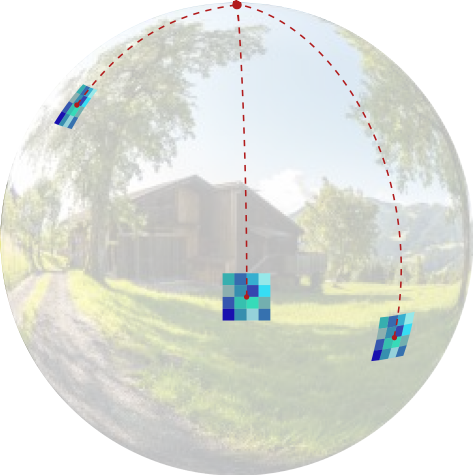}
\caption{In a \textit{coherent} filtering, the kernel is translated by keeping orientation with respect to the north pole, interpixel correlation and surface area.}
\label{fig:first_page} 
\end{figure}

A spherical image is a function defined over a sphere and the development of CNN processing tools over this 2D manifold is not straightforward. More precisely, the difficulty consists in constructing a convolution and its resulting CNN that satisfy the three following key properties: (i) \textit{rotation equivariance} (meaning that filtering and rotation commute), (ii) \textit{expressive} filter (the filter support can have any size and any impulse response), and (iii) \textit{computational efficiency} (complexity grows linearly with the number of pixels $N_{pix}$, and computation is local). Indeed, existing solutions maintain two out of these three characteristics. For instance, in a first set of methods, the sphere is mapped on one or multiple planes, and 2D-processing is applied on the plane(s) (equirectangular (ERP)~\cite{snyder1997flattening}, the cube map~\cite{ng2005data, kuzyakov2016next}, and the rhombic dodecahedron~\cite{fu2009the} projections). In these approaches, the convolution is not equivariant to rotation as the projection induces distortion, that depends on the position on the sphere. A second approach consists in performing CNN on a graph~\cite{perraudin_deepsphere_2019}. Here, the filters are restricted since only one coefficient per neighborhood can be learned, and not one per neighbor. In the last category of methods (multiple graph \cite{khasanova_geometry_2019}), and spherical harmonics based methods \cite{esteves_learning_2018,cohen_spherical_2018,esteves_spin-weighted_2020,roddy2021sifting}), the two first properties are maintained but the complexity is at least $N_{pix}^{3/2}$, see also \cref{tab:CNN_characteristics}. 


In this paper, by effectively integrating the geometry of the problem, we propose a novel learning approach for spherical images, called OSLO, that satisfies all above mentioned properties. First, to achieve rotation equivariance (i), we perform \textit{consistent} filtering. This means that, by translating the filter over the sphere surface, the filter (a) keeps the same orientation towards the north pole, (b) processes pixels with the same interpixel correlation, and (c) processes the same area of the sphere. These properties, illustrated in \cref{fig:first_page}, result from the exploitation of a sampling that is both uniform and oriented with respect to the north pole. It also takes advantage of the fact that omnidirectional images are usually level or registered\footnote{The north pole of these images usually point towards the upright direction such that buildings, humans, doors, etc. have a natural upright position.} \cite{uyttendaele_Optimizing_2017,davidson_360deg_2020}. Second, the image being registered, and the sampling being oriented, one coefficient per neighbor can be learned. Moreover, we propose an iterative construction of this convolution with a dedicated aggregation such that the kernel can be learned for every pixel neighborhood size. This leads to a filter with high expressiveness (ii). Finally, since filtering consists in translating the filter  over the sphere, complexity grows linearly with the number of pixels, and computation is local (iii).
 We also extend some of the important operators from classical 2D CNN to the sphere, such as stride, pooling, pixel shuffling, so that each elementary module of a 2D CNN can be defined on the sphere.

We finally illustrate the benefits of the proposed spherical framework, by extending two well-known, learning-based, 2D image compression architectures \cite{balle_end--end_2017,balle_variational_2018} onto the sphere. More precisely, we transform these two complete architectures directly on the sphere using all the elementary modules proposed in our OSLO solution. We then compare the compression performance of such architectures with the initial architectures from \cite{balle_end--end_2017,balle_variational_2018}, when applied to ERP images. The results show that our approach leads to significant compression gain with -18.3\% and -13.7\% of bit rate saving for the two architectures considered. Moreover, in the reconstructed images, the high frequency details are preserved at high bitrates, showing the benefits of the filter expressiveness. All this demonstrates the efficiency of the proposed convolution and CNN modules, and their potential impact for the extension of any architecture (performing any processing task) on the sphere.

The rest of the paper is organized as follows. Section \ref{sec:sota} explains in detail the challenges in designing a spherical CNN, and notably the difficulty to construct filters that satisfy all necessary properties for a CNN to be both efficient and tractable. We then present in Section \ref{sec:spherical_conv} our proposed 'On the Sphere Learning for Omnidirectional' (OSLO) framework and describe an efficient implementation. In Section \ref{sec:architecture}, we present an application of this new framework to omnidirectional image compression, and validate the benefit of our solution via extensive experiments. 

\section{Representation learning on the sphere}
\label{sec:sota}

\begin{table*}[]
\centering
\caption{Characteristics of spherical CNNs}
\label{tab:CNN_characteristics}
\resizebox{\textwidth}{!}{%
\begin{tabular}{ll|c|c|c|c||c|}
\cline{3-7}
                                                                                                                &                                            & \multicolumn{5}{c|}{Representation method}                                                                                        \\ \cline{3-7} 
                                                                                                                &                                            & 2D learning    & spectral learning & graph-based representation learning & multi-graph representation learning & proposed OSLO    \\ \cline{3-7} 
                                                                                                                & \multicolumn{1}{r|}{Domain of definition:} & mapped 2D grid & all               & all                                 & all                                 & uniform sampling \\ \hline \hline
\multicolumn{1}{|l}{\multirow{3}{*}{\begin{tabular}[c]{@{}l@{}}(i) filter\\ consistency\end{tabular}}}        & (a) same orientation                       & Yes            & Yes               & \textit{No}                         & Yes                                 & Yes              \\ \cline{2-7} 
\multicolumn{1}{|l}{}                                                                                          & (b) same interpixel correlation            & \textit{No}    & Yes               & Yes                                 & Yes                                 & Yes              \\ \cline{2-7} 
\multicolumn{1}{|l}{}                                                                                          & (c) same surface on the sphere             & \textit{No}    & Yes               & Yes                                 & Yes                                 & Yes              \\ \hline \hline
\multicolumn{1}{|l}{\multirow{2}{*}{\begin{tabular}[c]{@{}l@{}}(ii) filter\\ expressiveness\end{tabular}}}      & (a) anisotropic                            & Yes            & Yes               & \textit{No}                         & Yes                                 & Yes              \\ \cline{2-7} 
\multicolumn{1}{|l}{}                                                                                          & (b) large domain                           & Yes            & Yes               & Yes                                 & Yes                                 & Yes              \\ \hline \hline
\multicolumn{1}{|l}{\multirow{2}{*}{\begin{tabular}[c]{@{}l@{}}(iii) computational\\ efficiency\end{tabular}}} 
& (a) complexity                    
& $\mathcal{O}(N_{pix})$          
& at least  $\mathcal{O}(N_{pix}^{3/2})$       
& $\mathcal{O}(N_{pix})$                                 
& $\mathcal{O}(N_{pix}^{3/2})$                           
& $\mathcal{O}(N_{pix})$               \\ \cline{2-7} 
\multicolumn{1}{|l}{}                                                                                          & (b) local computation                      & Yes            & \textit{No}       & Yes                                 & Yes                                 & Yes              \\ \hline
\end{tabular}%
}
\end{table*}

\subsection{Problem settings}
The unit 2-sphere $\mathcal{S}^2$ is a 2D manifold defined as the set of 3D points in Euclidean space $\mathbb{R}^3$ with a norm of $1$. A spherical image $f$ is a continuous function $f: \mathcal{S}^2 \rightarrow \mathbb{R}^3$ representing red, green, and blue color channels\footnote{Here, we focus mainly on color~(RGB) images defined on the sphere since that is the most common use case, but all of our definitions are also valid to any multi-channel images, \emph{i.e.}, when $f: \mathcal{S}^2 \rightarrow \mathbb{R}^{\rm ch}$, with $\rm ch$ being the number of channels.}. For each point of the sphere $x \in \mathcal{S}^2$, the image value $f(x)$ corresponds to the color value of one light field's sample converging to the center of the sphere $O$ through $x$. In other words, it depicts the color of one visible point of the 3D scene lying on the line $(Ox)$. It is expected that $f(x)$ is correlated with $f(x')$ when $||x-x'||_2$ is small because they are likely to belong to the same object in the scene. In practice, the spherical image is captured with a specific hardware that depends on the application. In general, it comes from the combination of several captures that can be done with perspective, fish eye, or catadioptric cameras \cite{yagi1999omnidirectional}. Then, in most applications, these images are not rendered in their spherical shape. For example, in virtual reality applications, they are partly projected on a Head Mounted Display~(HMD) that is controlled by a user. 
In such a context, it is preferred that the images are level or registered. In other words, the north pole of these images should point towards the upright direction such that buildings, humans, doors, etc. have a natural upright position. A non-level 360$^\circ$ image completely breaks the sense of realism for virtual reality users and leads to an unpleasant immersive experience and even severe user sickness \cite{padmanaban2018towards}. In case the 360$^\circ$ images are not upright, they can be leveled using \cite{uyttendaele_Optimizing_2017,davidson_360deg_2020}. Another consequence of the rendering modes is that the images should be of a very high resolution (up to 10K) so that the HMD visualization remains of a sufficiently good quality.

Like 2D images, spherical images are employed in various contexts for which efficient processing tools are required: compression, object recognition, classification, super-resolution, etc. Such tasks may benefit from the recent advances in machine learning, and therefore, there is nowadays a crucial need for defining proper representations for spherical data. However, due to their special domain of definition, huge size, and specific geometry distortion, spherical images are not straightforwardly processed by traditional CNNs. Specific spherical CNN architectures have to be constructed while respecting a certain number of requirements summarized in \cref{tab:CNN_characteristics}. 
First, the convolutional filters have to be \textit{rotation equivariant}, \emph{i.e.}, filtering and rotation commute. This enables the filters involved in the CNN architecture to be \emph{applied consistently all over the sphere} (see \cref{fig:first_page}). Meaning that the learned filters should keep the same orientation with respect to the north pole and surface area on the sphere, while modeling the same inter-pixel correlation. 
Second, to guarantee a good processing performance, the filters used at the different layers should be \emph{as expressive as those of the common 2D CNN architectures}. Meaning that they should be able to perform convolution on a \emph{large neighborhood}, and have \emph{anisotropic} responses. Finally, the filters computation should be \emph{efficient}. In other words, the computation of the convolution should be linear with the number of pixels (as in a standard 2D CNN), and a local computation should be feasible (not involving the whole image at each filter).

In the next section, we will discuss the challenges of achieving these goals for spherical data.


\subsection{Challenges}

On-the-sphere operations such as convolution and translation admit several definitions. Thus, filtering a spherical image by a kernel can be performed in several manners, each of them raising specific issues. 

A first definition represents the sphere with SO(3), \emph{i.e.}, the group of rotations in $\mathbb{R}^3$, and computes a filtered output of a spherical image for every rotation in SO(3), \emph{i.e.}, moving the kernel on the sphere with 3 degrees of freedom. While exhaustive, this filter representation has the drawback of an increased dimension~(from 2 to 3) during the filtering process.

Another definition restricts the displacement of the kernel to 2 degrees of freedom. In such a case, the spherical convolution is defined as the inner product between the image $f$ and a localized kernel $h$ translated at different positions $x \in \mathcal{S}^2$:
\begin{equation}
    (f \star h)(x) = \langle f, \mathcal{T}_x h \rangle, \label{eq:siftingconv}
\end{equation}
where $\mathcal{T}_x$ is a translation operator. We can see that the filtering of a spherical image by a kernel remains a spherical image defined on $\mathcal{S}^2$. However, there are an infinite number of ways of defining $\mathcal{T}_x$, depending, for example, on the direction of the polar axis. This ambiguity may be problematic in some applications, in particular when the spherical data does not have any a priori fixed orientation. However, this definition can be advantageous when dealing with registered (upright) omnidirectional images since it ensures that the kernel is always oriented similarly and consistently with the south/north axis at every position (as illustrated in \cref{fig:first_page}). 


Even with only 2 degrees of freedom, the definition of the translation $\mathcal{T}_x$ is challenging. It can even become an open problem when dealing with discrete pixel positions.
On top of the necessary interpolation artifacts, the pixelization of the sphere impacts the structure and properties of the discrete domain of definition.
On the one hand, as 2D translations are well defined on a Cartesian grid, one can map the sphere onto one or multiple 2D planes. This unfortunately results in heterogeneous (and thus inconsistent) pixel distributions on the sphere, which have to be compensated at the filtering stage. This is not straightforward, and may lead to complex operations or necessary simplifications of the filters, penalizing at the same time their expressiveness.
On the other hand, when the sphere is pseudo-uniformly sampled, the pixel's set becomes non-Cartesian, and therefore $\mathcal{T}_x$ is undefined.
In a nutshell, we can see that the main challenge raised by spherical learning is to find the adequacy between the sphere sampling, the signal processing tools that it enables, and the compensation of their limitations.


\subsection{Current Solutions}

We now review in detail the solutions proposed in the literature to learn representations for omnidirectional images. 
Their properties in terms of processing accuracy and efficiency are summarized in \cref{tab:CNN_characteristics} and discussed in detail in the following. 

\subsubsection{2D learning}
Mapping spherical images on 2D planar grid enables to directly apply the well-known 2D CNN tools on such a planar representation. For instance, authors in \cite{su_learning_2017} use the popular equirectangular projection~(ERP) that samples the sphere with fixed longitude and latitude steps. ERP sampling, however, results in highly heterogeneous pixel distribution on the sphere.
As a consequence, they cannot guarantee that the kernel keeps the same surface nor describes a constant inter-pixel correlation over the sphere (see \cref{tab:CNN_characteristics}).
To circumvent this issue, the authors adapt the size of the kernel at different elevations by learning different variable-size kernels for each row of the ERP. However, this dramatically increases the model size and does not fully utilize CNN's weight-sharing property.

Instead of ERP, the works in \cite{monroy_SalNet360_2018,ruder_artistic_2018} propose to use cube map projection~(CMP). The image is mapped to the six faces of a cube, and each face is processed with conventional 2D CNN, considering the face as the image plane of a perspective camera. 
While the CMP pixel distribution on the sphere is more uniform than ERP's one, the discontinuities at the borders of the cube faces require to be handled carefully that needs post-processing. Also, defining a kernel that keeps its orientation constant with respect to the north pole is impossible since either the north pole lies in the middle of the top face, or worst at one of the corners of the cube.

Other mappings have been investigated in~\cite{lee2019spherephd,zhang2019orientation,cohen2019gauge}. By increasing the number of faces of a polyhedron on which the sphere is projected, the pixel's distribution on the sphere is closer to a uniform pixelization. However, the border discontinuity problem is increased, and the orientation of the filter with respect to the north is not always guaranteed. Despite their convenient link with existing 2D tools, mapping-based solutions inherently lead to too many drawbacks, (\emph{e.g.}, non-uniform sampling, discontinuities and local orientation loss) to be convincing. 

\subsubsection{Spectral learning}
Similar to classical 2D convolution, one can define convolution on the sphere as a product in the Fourier domain. 
Such an idea is interesting since the spherical harmonics are a good Fourier basis to represent spherical signals. 
More precisely, as summarized in \cref{tab:CNN_characteristics}, they enable defining consistent and expressive filtering all over the sphere.
However, these methods are complex, and the simplifications proposed to reduce the computation time often imply a loss in expressiveness. 
For instance,  Esteves et al.~\cite{esteves_learning_2018} propose a spherical convolution with reasonable complexity $\mathcal {O}(N_{pix} \log{}^2 N_{pix})$, but it only generates isotropic filters. 
Instead, Cohen et al.~\cite{cohen_spherical_2018} propose using spherical correlation defined in SO(3).
Such an approach enables the convolution parameters to take into account filter direction.
Nevertheless, the complexity of this approach~($\mathcal {O}(N^2_{pix})$) rapidly becomes a limitation when handling high-resolution spherical images. 
Recently, new spectral spherical CNNs with spin-weighted functions have been proposed~\cite{esteves_spin-weighted_2020}.
The proposed convolution is not restricted to isotropic filters and remains 2D and defined on the sphere.
However, the $\mathcal {O}(N^{3/2}_{pix})$ complexity still limits its use to high-resolution images. 
Similarly, Roddy et al.~\cite{roddy2021sifting} have proposed a sifting convolution defined as the inner product between the image $f$ and a localized kernel $h$ translated at different positions in the spherical harmonics domain~(with 2 degrees of freedom). This approach has strong similarities with dictionary-based ones~\cite{tosic2006progressive}. Indeed, atoms are defined locally and translated on the sphere to form a dictionary used for image representation.
While such approaches match the consistency and expressiveness expectations summarized in \cref{tab:CNN_characteristics}, their supra-linear complexity remains a major limitation for high-resolution spherical images. Moreover, local estimation of the filtering process is not possible since spherical harmonics have to be estimated on the whole sphere. 

\subsubsection{Graph-based representation learning}

Recently, graph signal processing tools have demonstrated their efficiency when dealing with non-euclidean topology. By modeling the sphere as a graph and embedding the inter-pixel distance in the edge weights, a filter can be applied consistently all over the spherical data, independently of the chosen sampling. For example, Khasanova and Frossard~\cite{khasanova_graph-based_2017} propose a graph-based convolution for ERP in which the edge weights between neighboring nodes~(pixels) are equal to the inverse of their distances.
Alternatively, other works use a quasi-uniform sampling of the sphere and define graph-based CNNs directly on the sphere~\cite{perraudin_deepsphere_2019,yang_rotation_2020}.
Perraudin et al. \cite{perraudin_deepsphere_2019} use Hierarchical Equal Area isoLatitude Pixelation~(HEALPix) \cite{Gorski_2005}, and Yang et al.~\cite{yang_rotation_2020} use Geodesic ICOsahedral Pixelation (GICOPix). To keep a linear complexity filtering, the convolution is approximated using a Chebyshev polynomial formulation, thus avoiding the graph spectral transform computation.
Despite their great adaptivity to the topology and a good filtering consistency over the sphere, the main drawback of such methods is the low expressiveness of the learned filters (see \cref{tab:CNN_characteristics}. Due to the lack of direction in the graph, only isotropic filters can be learned, which is a major issue in imaging applications.

\subsubsection{Multi-graph representation learning}

To overcome the aforementioned limitations of graph-based approaches, Khasanova and Frossard~\cite{khasanova_geometry_2019} use multiple directed graphs, in which each graph represents a specific direction and location of neighboring pixels relative to the central pixels. Such an approach allows adaptation to any topology while preserving orientation information and filter expressiveness.
However, their graph construction is complex and requires a large storage space. Indeed, to create an $n \times n$ convolution kernel, $n^2$ graphs are constructed \textit{at each pixel position}. 
More precisely, the graph $k$ is constructed by projecting the k\textsuperscript{th} area of an $n \times n$ tangent rectangle on the sphere surface and linking all pixels in the projected area to the central pixel. This implies a high complexity to estimate all the $n^2$ graphs at each kernel position and requires a large storage size~(for each layer of the neural network).

\subsection{Discussion}
As we can see in \cref{tab:CNN_characteristics}, none of the current learning approaches is able to satisfy all the initial requirements and have at the same time a linear complexity, \emph{i.e.}, $\mathcal{O}(N_{pix}$). While it is not possible to guarantee consistent filtering with the mapping-based solution, using graph-based approaches to adjust to the topology reduces the filter expressiveness. Spectral and multi-graph approaches support more expressive filters that can be applied consistently. However, they come with a significant increase in complexity, resulting in issues to handle high-resolution 360$^\circ$ images. Alternatives to the aforementioned categories of methods have also been explored by defining tools directly on the uniformly sampled sphere~(\emph{e.g.}, \cite{jiang2018spherical,krachmalnicoff_convolutional_2019}). Their approaches however remain pretty complex and too specific to the target applications.


\section{On-the-Sphere Learning for Omnidirectional image} \label{sec:spherical_conv}



\subsection{OSLO Overview}

As discussed earlier (\cref{tab:CNN_characteristics}), existing spherical learning methods fail to construct expressive and consistent filters while being computationally tractable. Thus, we propose a new framework, On-the-Sphere Learning for Omnidirectional Images (OSLO), to build effective representations of spherical data that attain the same filter expressiveness, consistency, and complexity of 2D CNNs.  More precisely, we rely on the HEALpix sampling's properties to redefine all the elementary modules composing a CNN architecture \emph{on-the-sphere}. In addition to the convolution and pooling operations that appear in most deep network models, we define 2D CNN's operators such as stride, skip connection, and patching \emph{on-the-sphere}. In this section, we present the design of the different operators, which can eventually be assembled to create different architectures trained on diverse tasks. One example of using OSLO for neural compression of omnidirectional images is presented in \cref{sec:architecture}. 

\subsection{Sampling}


\begin{figure} 
    \centering
  \subfloat[\label{subfig:regularity}]{\fontsize{7pt}{6pt}\selectfont
\def\svgwidth{0.45\linewidth}
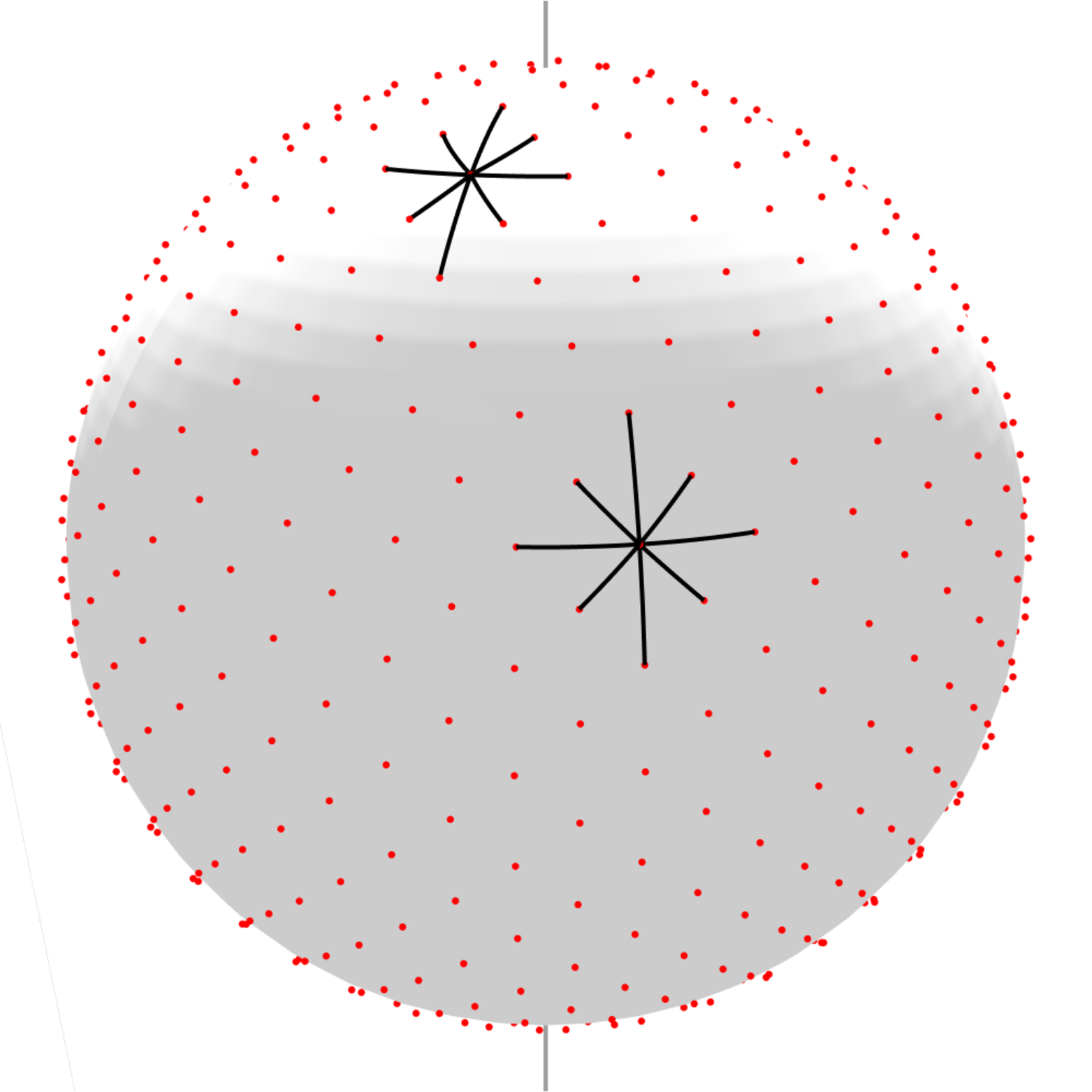}
    \hfill
  \subfloat[\label{subfig:rigidity}]{\fontsize{7pt}{6pt}\selectfont
\def\svgwidth{0.45\linewidth}
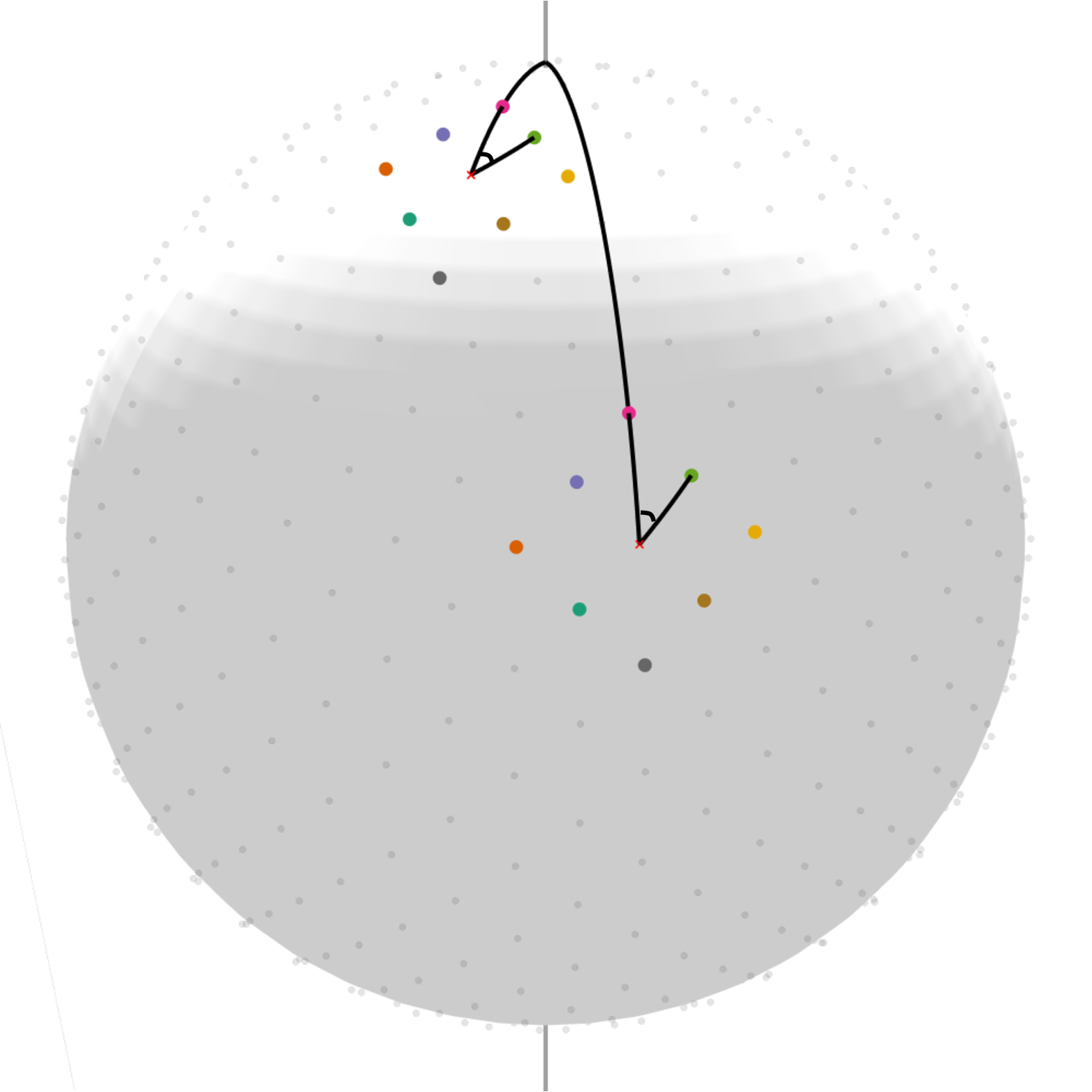}
  \caption{Desired sampling's properties : (i) It should be \textit{regular} such that each pixel has same number of neighbors, and we should be able to discriminate between neighboring pixels by assigning different directions to each neighbor (as shown in (a)); (ii) The sampling should be \textit{rigid} (as shown in (b)) such that the distance between a specific neighbor (1 to 8) and the center pixel should be almost constant all over the sphere (\textit{e.g.}, $d_1=d_1^{\prime}$). Additionally, its relative orientation with respect to the center pixel should be constant over the sphere, (\textit{e.g.}, $\theta_1=\theta_1^{\prime}$). This orientation is computed with respect to the north pole.}
  \label{fig:regularity_rigidity} 
\end{figure}

To be able to have fast convolution and avoid using spectral domain, we propose implementing the spherical convolution in the pixel domain, which corresponds to translating a kernel as defined in \eqref{eq:siftingconv}.
One of the key elements in defining an effective representation learning solution is the choice of a proper sampling method that enables the design of expressive filters with a fast implementation. The filter output is computed with the same weights, irrespective of the position of the output on the sphere. As a consequence, the complexity scales linearly with the number of pixels.
In order to construct this computationally efficient convolution, the chosen sampling must fulfill some properties summarized in \cref{fig:regularity_rigidity}.
First, to have an expressive filter, the sampling grid must ensure \emph{regularity} of the neighborhood such that the same number of neighbors always surrounds each pixel, and those can be identified~(based on their relative position with respect to the central point). 
Second, the sampling grid should ensure that the relative distance and orientation between a point and its neighbor is constant over the sphere; thus, the \emph{rigidity} of the neighborhood and, therefore, the consistency of the filter shape is guaranteed. 
Several sampling methods exist, most of them relying on a sphere tessellation. Often, only one of the regularity or rigidity constraints is achieved. For example, being pseudo-uniform as GICOPIX \cite{yang_rotation_2020} is not sufficient since the pixels have additionally to be organized such that the translation of the kernel remains coherent over the sphere. In other words, a specific neighbor that has a specific position relative to the center pixel must be easily identifiable throughout the sphere.

We now show that HEALPix \cite{Gorski_2005} is a reliable sampling process for our convolution that fulfills the above requirements for effective and expressive feature representation. In HEALPix, the tessellation process begins by partitioning the spherical surface into 12 equal-area regions~(base resolution). To have finer pixelization, each region is recursively divided into 2$\times$2 equal-area sub-pixels until the desired resolution is reached.
By construction, each pixel in HEALPix has eight adjacent neighbors~(structured in a diamond pattern), except for 24 pixels that have seven neighbors~(for any resolution higher than the base resolution). The effect of these exceptional points is negligible for high-resolution 360$^\circ$ images~(a HEALPix resolution close to conventional 4K resolution has less than 0.0002\% pixels with 7 neighbors). Finally, a fundamental property of HEALPix, often overlooked in the literature, is that the orientation of neighboring points with the central point is almost fixed all over the sphere. Each neighbor can be identified by its relative direction to the central pixel: SW, W, NW, N, NE, E, SE, and S. Therefore, HEALPix fulfills the \emph{regularity} property that is essential to define an expressive filter. 

\begin{figure} 
    \centering
      \subfloat[\label{subfig:equi_neighboring}]{\includegraphics[width=0.45\linewidth]{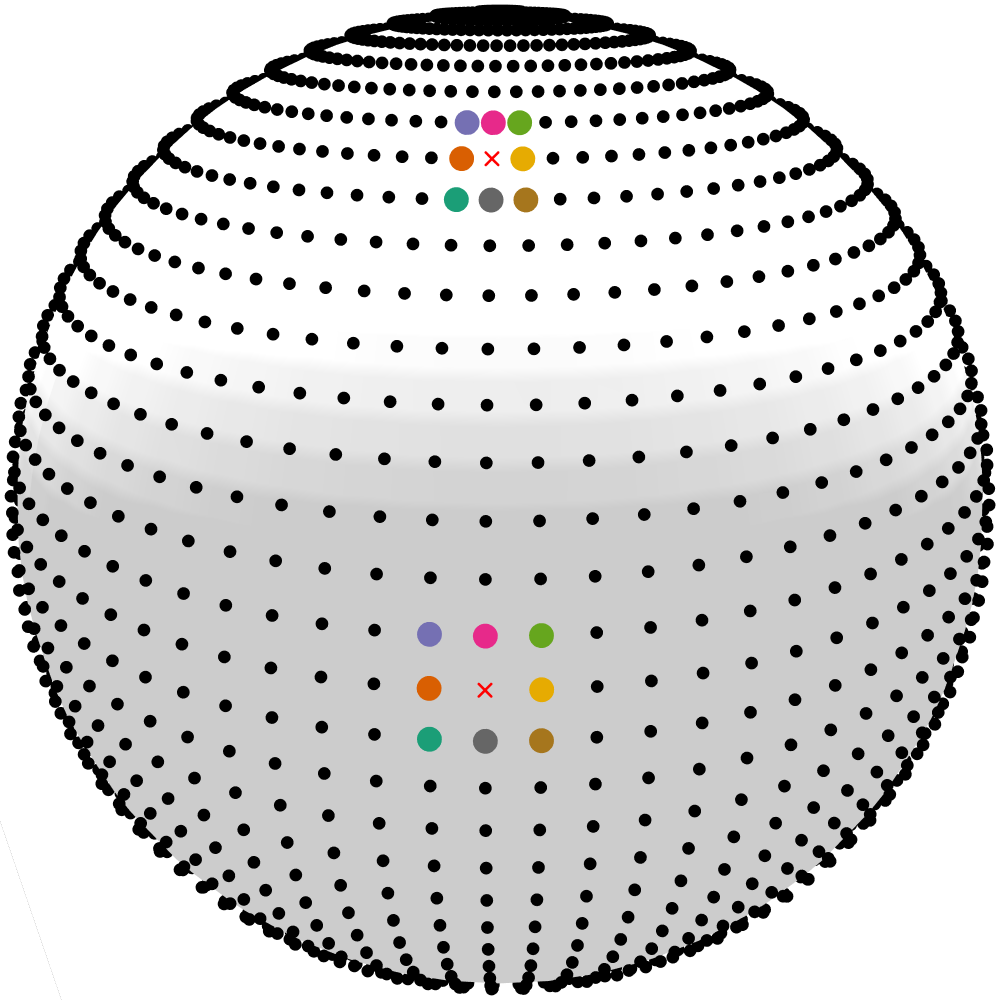}}
    \hfill
  \subfloat[\label{subfig:healpix_neighboring}]{\includegraphics[width=0.45\linewidth]{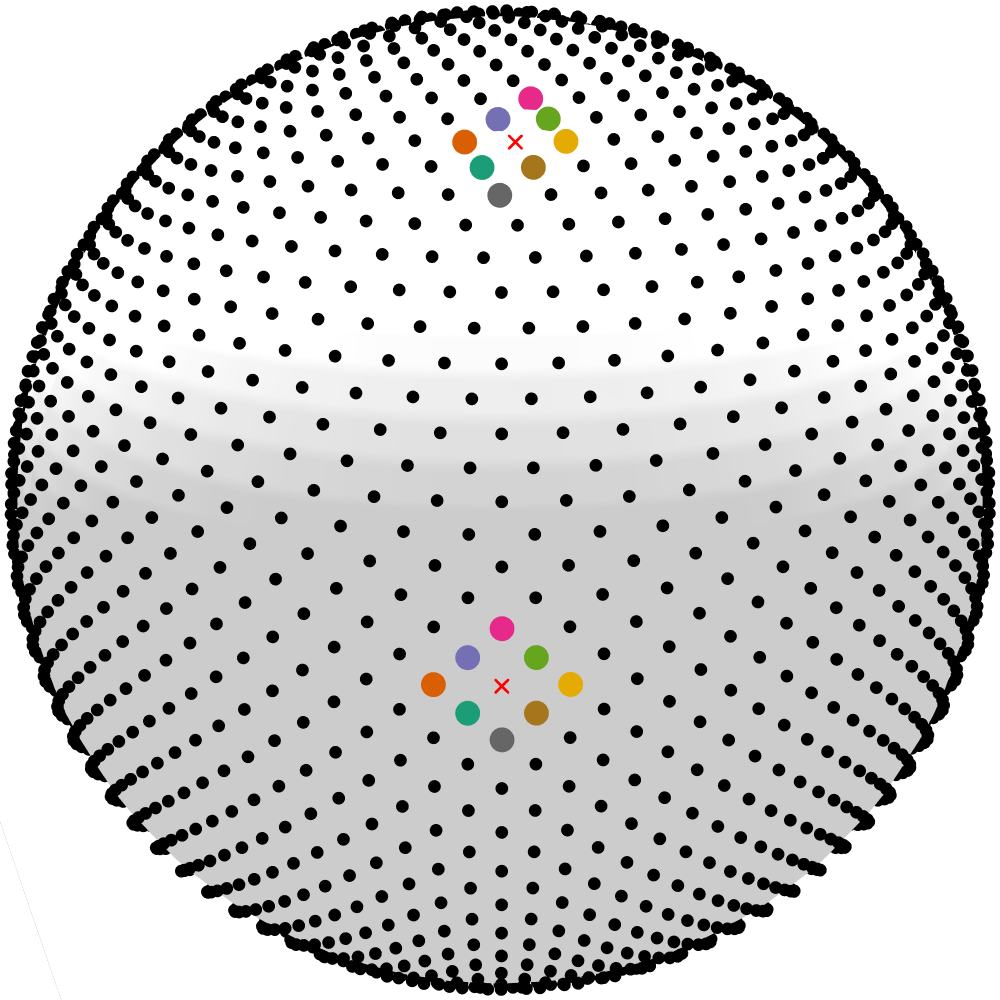}}
  
  \subfloat[\label{subfig:equi_neighboring}]{\includegraphics[width=0.45\linewidth]{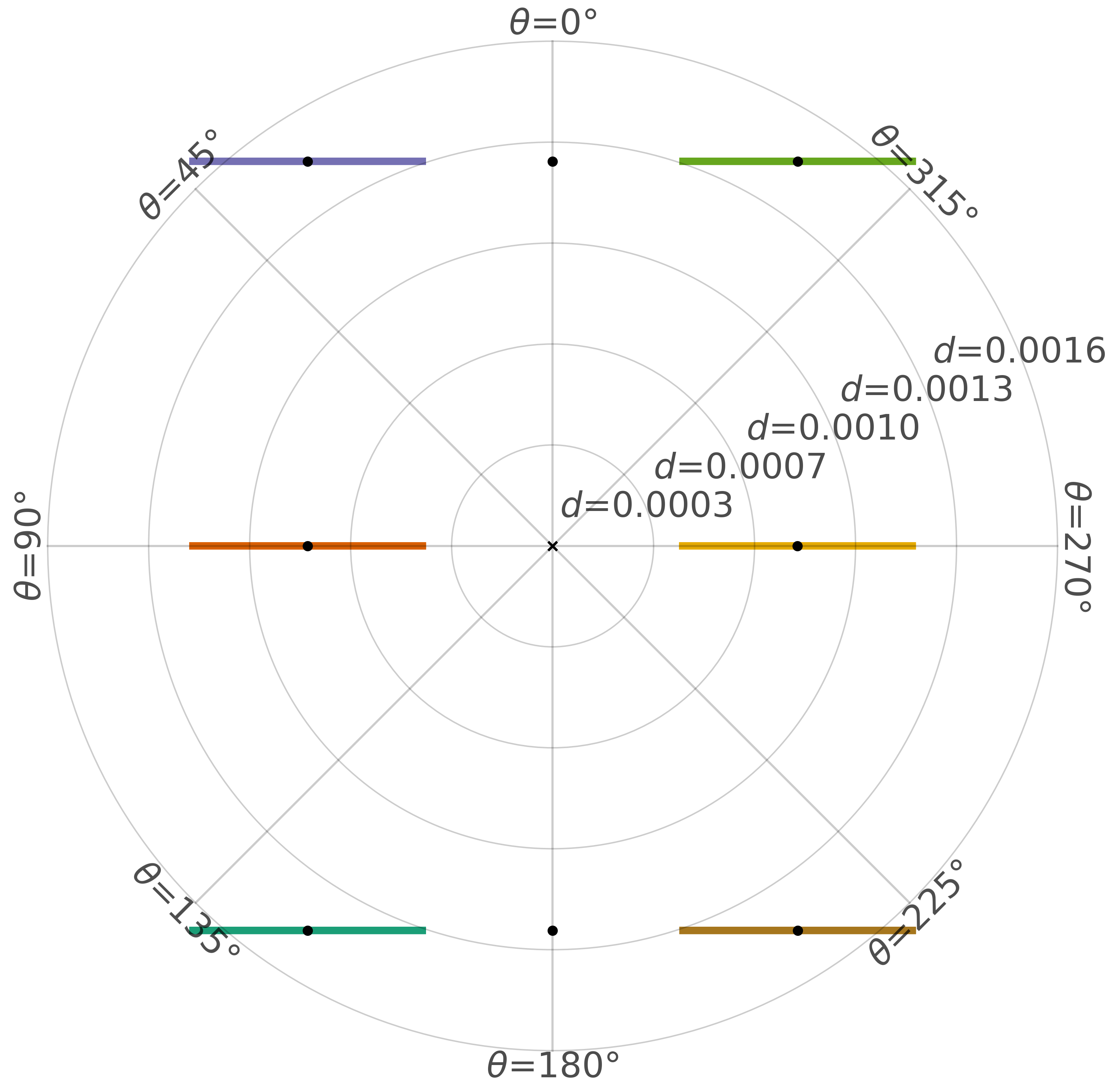}}
    \hfill
  \subfloat[\label{subfig:healpix_neighboring}]{\includegraphics[width=0.45\linewidth]{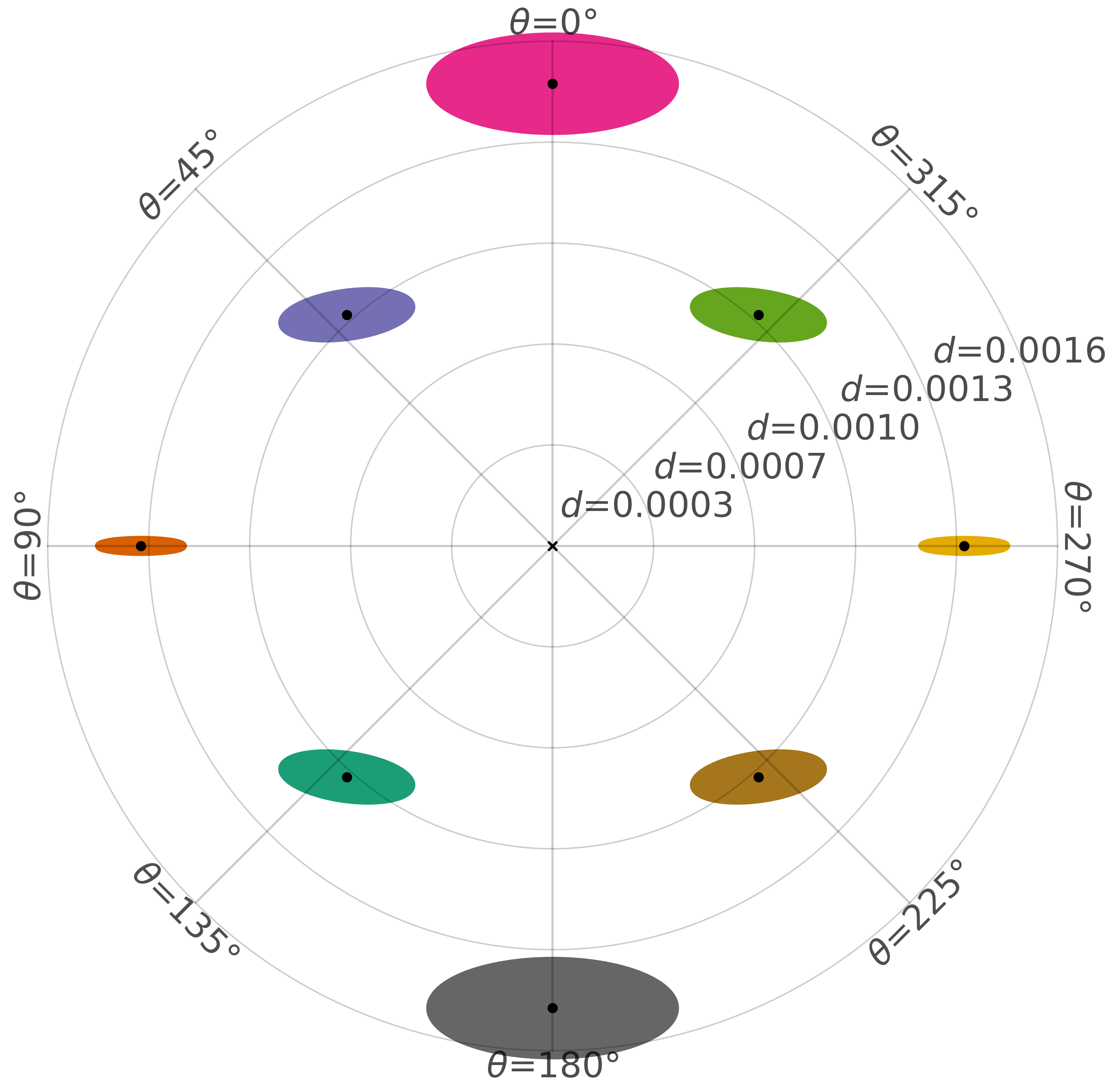}}

  \caption{The distribution of neighboring pixels around the central pixel. Left column: ERP. Right column: HEALPix sampling. Top row: Pixel positions on the sphere when ERP and HEALPix have almost the same number of pixels with 3072 and 39$\times$78 pixels. Bottom row: Pixel distributions on the tangent plane when ERP and HEALPix have almost the same number of pixels with 12582912 and 2508$\times$5016 pixels. Marker $\times$ shows the central pixel, black circles represent the mean position of each neighbor. Each color corresponds to a neighbor pixel in a specific direction (SW, W, NW, N, NE, E, SE, and S) and the corresponding area's size is set based on the standard deviation of pixel distribution in the plane tangent to the sphere at the central pixel.}
  \label{fig:neighboring_distribution} 
\end{figure}

To demonstrate the \textit{rigidity} property of HEALPix, we calculate the relative distance and angle ($d$ and $\theta$ of \cref{fig:regularity_rigidity}) for each position on the sphere. More precisely, for a given central pixel, we define a tangent plane in which the y axis points towards the north pole. Then, each neighboring pixel is projected onto the tangent plane. The relative distance and angle to the central pixel are then computed. By doing that for every pixel position on the sphere, we are able to show the neighbor's distribution. The distribution of neighboring points for both HEALPix and ERP are shown in \cref{fig:neighboring_distribution}. 

At first sight, it is clear that HEALPix pseudo-uniform sampling provides a \emph{rigid} structure, allowing each neighbor to keep its relative distance and direction with respect to the center point, all over the sphere. When compared to ERP, for the majority of neighbors (6 out of 8 neighbors), HEALPix has much lower discrepancies. North and south neighbors present greater discrepancies than ERP. However, since the correlation in 360$^\circ$ images is mostly horizontal \cite{maugey_Saliency-based_2017}, the north and south neighbors have less impact than other pixels. Furthermore, the discrepancy in HEALPix does not affect the distance to the central node in comparison to the ERP. The relative standard deviation (ratio of standard deviation to the mean) of the distances between each neighbor and central pixel is given in \cref{tab:neighboring_distance}. This is important for the design of a convolution filter because the signal correlation depends on the distance of central pixels to their neighbors \cite{khasanova_graph-based_2017,perraudin_deepsphere_2019}. In other words, the impact of kernel weights on neighboring pixels is affected by their distance to the central node, and the filter response is effective if this distance is uniform throughout the sphere.

Since HEALPix fulfills the regularity and rigidity properties, we propose to use that sampling method to develop computationally efficient operators on the sphere. In particular, we can now define a new linear complexity convolution operator, and show that it leads to expressive and consistent filters (properties (i) and (ii) in \cref{tab:CNN_characteristics}).

\begin{table}[]
\centering
\caption{Relative standard deviation (in \%) of distances with respect to the central pixel for each neighbor all over the sphere. Both samplings have the same number of pixels (HEALPix sampling with 12582912 pixels and ERP of size 2508$\times$5016)}
\label{tab:neighboring_distance}
\resizebox{\linewidth}{!}{%
\begin{tabular}{l|c|c|c|c|c|c|c|c|c|}
\cline{2-10}
                              & SW             & W              & NW             & N          & NE             & E              & SE             & S          & Mean           \\ \hline
\multicolumn{1}{|l|}{ERP} & 12.01          & 48.34          & 12.01          & \textbf{0} & 12.01          & 48.34          & 12.01          & \textbf{0} & 18.09          \\ \hline
\multicolumn{1}{|l|}{HEALPix}    & \textbf{11.21} & \textbf{10.28} & \textbf{11.21} & 13.90      & \textbf{11.21} & \textbf{10.28} & \textbf{11.21} & 13.9       & \textbf{11.65} \\ \hline
\end{tabular}%
}
\end{table}

\subsection{Spherical convolution}
\begin{figure} 
    \centering

    \begin{minipage}[b]{.48\linewidth}
      \subfloat[\label{subfig:kernel_on_sphere}]{\includegraphics[width=\linewidth]{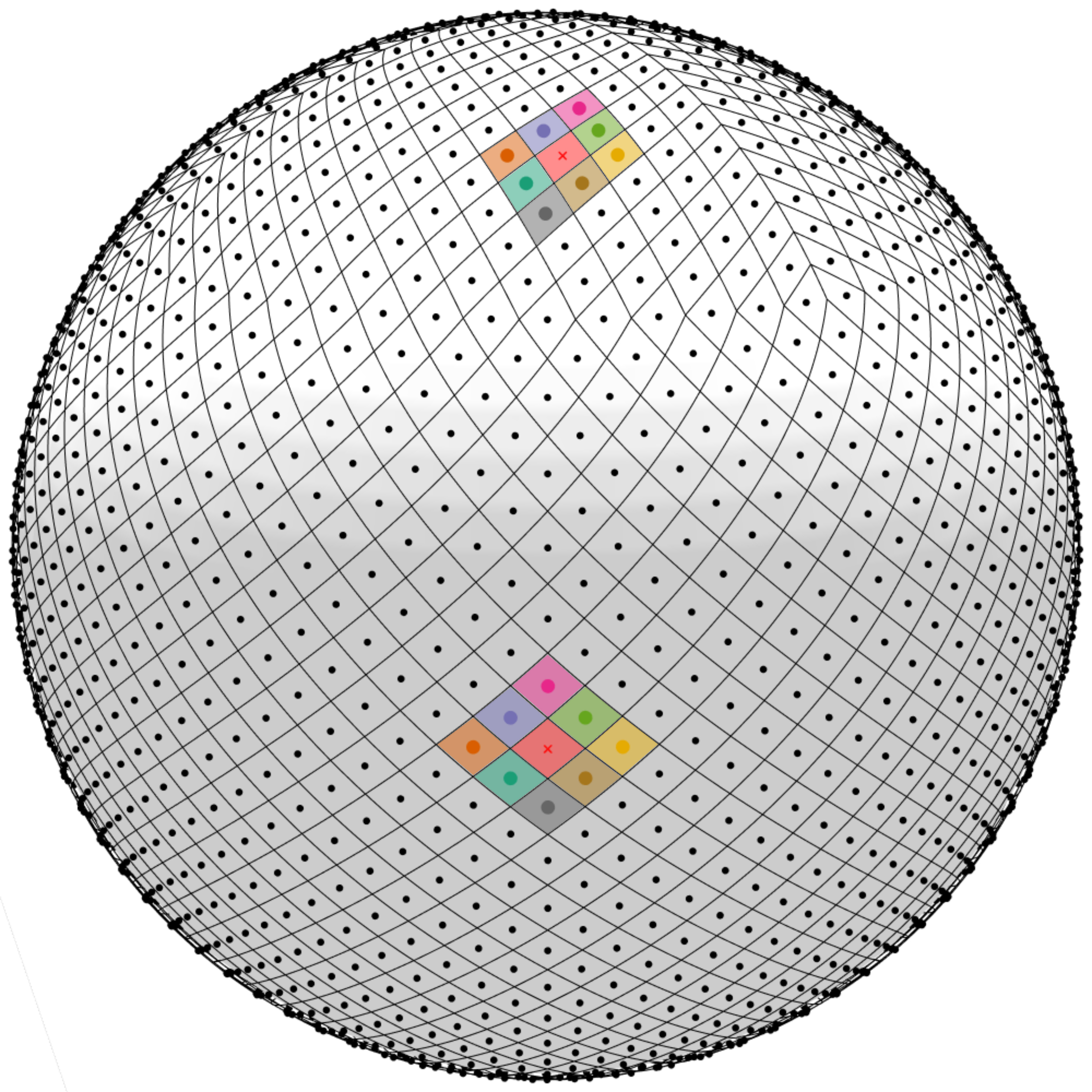}}
    \end{minipage}
    \hfill
    \begin{minipage}[b]{.48\linewidth}
    \centering
	    \subfloat[\label{subfig:message_passing_conv}]{\includegraphics[width=\linewidth]{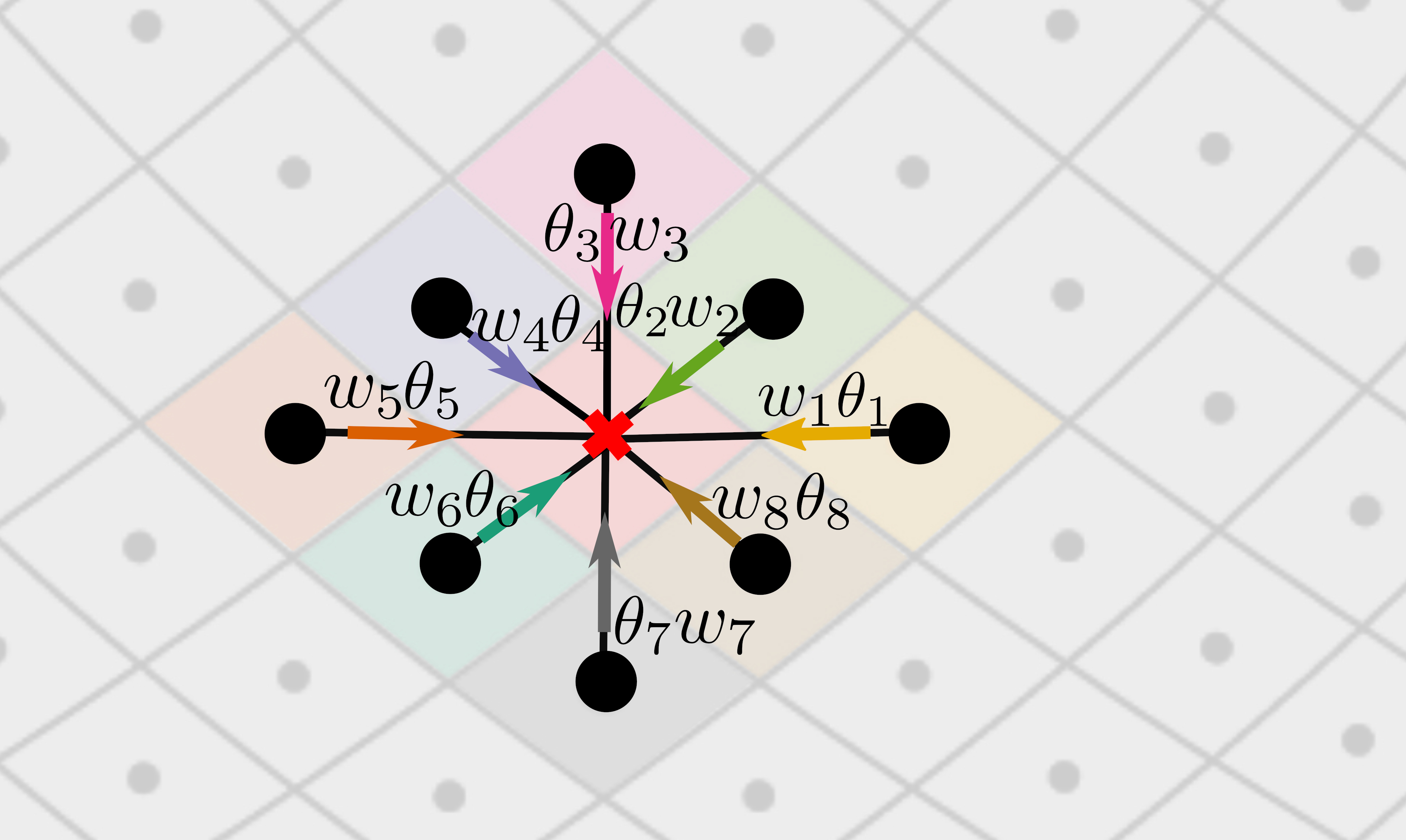}}
	\vfill
	    \subfloat[\label{subfig:pixel_pair_contribute_differently}]{\includegraphics[width=\linewidth]{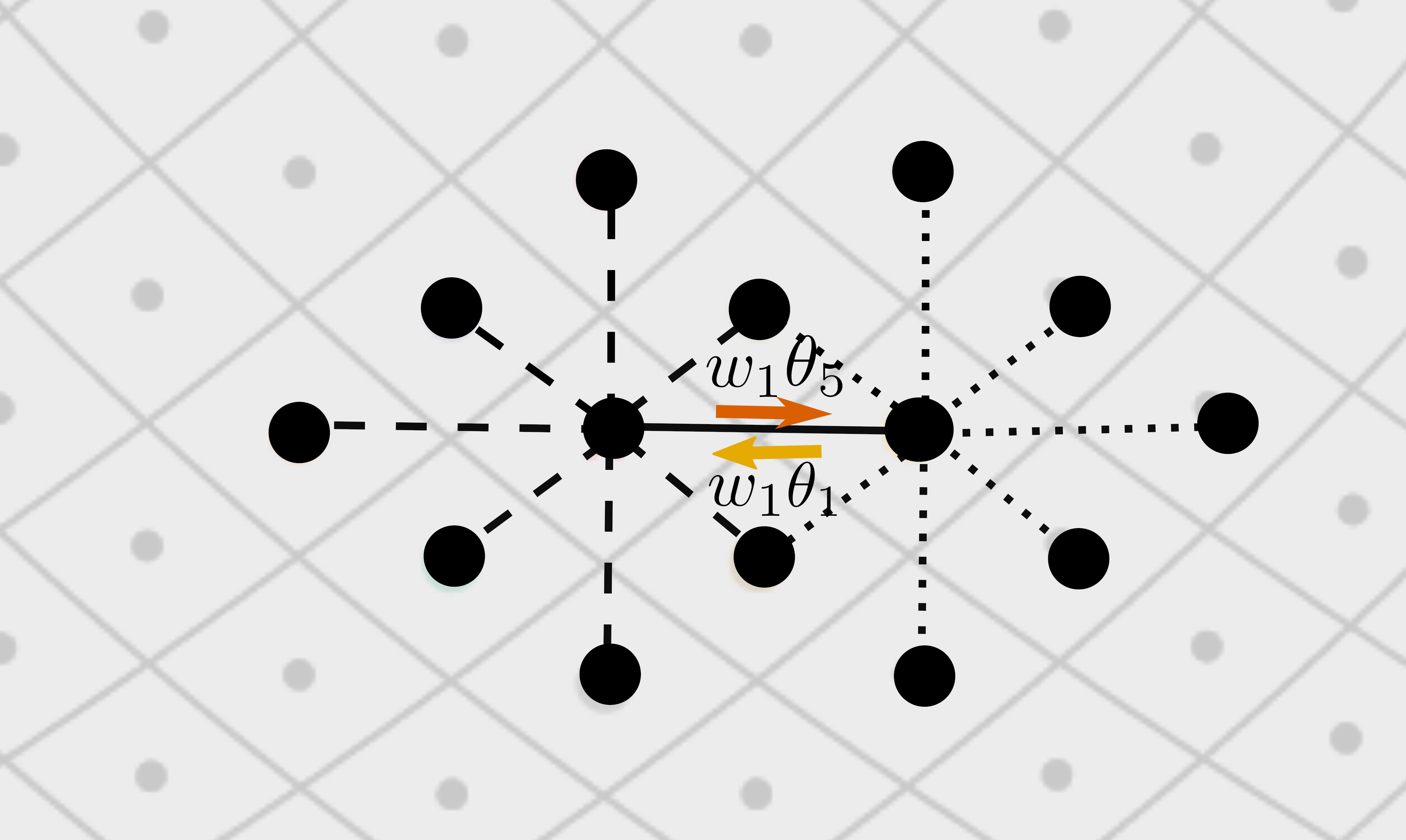}}
	\end{minipage}

  \caption{Our spherical convolution. (a) Kernel on the sphere at different positions. Each color corresponds to a different learnable kernel weight corresponding to a different orientation. (b) Convolution operation can be seen as a message-passing scheme. (c) The same pair of pixels contribute differently depending on the position of central pixel.}
  \label{fig:spherical_convolution} 
\end{figure}

Since we can distinguish the orientation of neighbors relative to the center and the neighborhood structure is fixed all over the sphere, it is possible to assign different learning weights to different directions, as shown in \cref{subfig:kernel_on_sphere}. Let $\mathcal{N}_i(k)$ denote the index of $k$-th neighbor of vertex $i$ ($k=1 \cdots 8$ representing \{SW, W, NW, N, NE, E, SE, S\}), and let $L_{in}$ and $L_{out}$ be the number of input and output features in the convolution, respectively. For each output feature $l, 1\le l \le L_{out}$, we define the convolution operation as: 
\begin{equation}\label{eq:conv}
x^{l}_i = \langle \mathbf{\Theta}_0, \mathbf{x}_i \rangle+ \sum_{k=1}^{8} \langle \mathbf{\Theta}_k, \mathbf{x}_{\mathcal{N}_i(k)} \rangle \cdot w_{\mathcal{N}_i(k), i}\text{ ,}
\end{equation}
where $\mathbf{\Theta}_k$ is the learnable weights of the CNN filters, $\mathbf{x}_i$ represents the input signal/features at vertex $i$, and $\langle \cdot, \cdot \rangle$ denotes the inner product operator. Both $\mathbf{\Theta}_k$ and $\mathbf{x}_i$ are $L_{in}$-length vectors. To handle the 24 exceptions of HEALPix pixels that do not have $8$ neighbors, we define:
\[
    w_{\mathcal{N}_i(k), i} = 
\begin{cases}
    0,& \text{if the neighbor $\mathcal{N}_i(k)$ is missing}\\
    1,              & \text{otherwise}
\end{cases}
\]

Interestingly, the proposed filter is anisotropic and consistent all over the sphere~(properties (i.a) and (ii) in \cref{tab:CNN_characteristics}). Note also that our proposed convolution can be interpreted as an update rule of a message-passing algorithm applied to graphs \cite{gilmer_2017_neural}. Such an interpretation is of great importance since it allows us to define convolution with larger filter sizes and therefore further gains in terms of filter expressiveness. The proposed convolution, however, is a generalization of the classical update rule in graph message-passing since the weighting applied to a message depends on the direction of the edge, as illustrated in \cref{subfig:pixel_pair_contribute_differently}.

\subsection{Increasing the local support to $n$-hop neighborhood}\label{subsec:extension_hop}

Extending the support of the filters is of great importance in order to increase the filter expressiveness (\emph{i.e.}, property (i) in \cref{tab:CNN_characteristics}). This requires generalizing the filtering operation \eqref{eq:conv} to $n$-hop neighborhoods and, therefore, being able to identify the nodes in these larger neighborhoods. However, direct identification of the neighbors and their relative positions to the central pixel is a complex operation and requires a large amount of storage space that grows quadratically with $n$. More precisely, for $n$-hop neighbors we need to store the index of $(2n+1)^2-1$ neighbors per pixel and per layer with different resolutions. In this section, we propose a fast and efficient iterative convolution that permits to automatically compute the convolution for any $n$-hop neighborhood. Our solution relies on the aggregation of $n$ $1$-hop neighborhood convolutions. This allows us to approximate the heavy convolution of a large neighborhood by iterative operation of lighter convolutions that only require access to direct neighbors. Although the expressiveness, in terms of the number of learned parameters, may be decreased with such a strategy, the resulting filter remains directional and quite complex, especially for small $n$. Moreover, in practice, this strategy has obtained promising results in the context of 2D CNN and graph-based CNN~\cite{he_deep_2016,xu_representation_2018}.

\begin{figure}
\centering
\fontsize{7pt}{6pt}\selectfont
\def\svgwidth{1\linewidth}
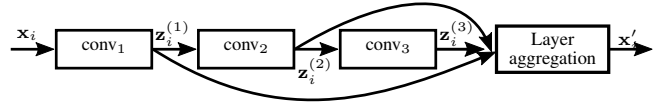
\caption{Proposed strategy to increase the local support (here $3$-hop) in spherical convolution. Each conv$_l, l=1,2,3$ performs neighborhood aggregation for each pixel as given in \eqref{eq:conv}. Finally, the last layer aggregation module combines pixel-wise the generated feature maps of the intermediate layers.}
  \label{fig:hop_extension} 
\end{figure}

Theoretically, a cascade of $n$ neighborhood aggregation makes use of a subtree structure of depth $n$ rooted at every pixel. Although deeper versions of neighborhood aggregation have access to more information, they may not provide better performance as the local information is diluted by too much averaging through successive neighborhood aggregation. Additionally, with deep neighborhood aggregation, vanishing gradient problem is encountered in the backpropagation of the training phase. For example, it has been observed that the best performance with graph convolutional networks is achieved with a cascade of two 1-hop convolutions~\cite{xu_representation_2018}. To overcome this limitation, skip connections are utilized in computer vision and graph learning \cite{xu_representation_2018} where the output of each iteration is sent to a further layer. Finally, a pixel-wise aggregation mechanism combines the features of intermediate layers independently. Formally, let $\mathbf{z}_i^{(l)}, l=1 \cdots n$ denote the hidden feature of pixel $i$ at layer $l$ that is sent to the last layer as shown in \cref{fig:hop_extension}. A skip strategy is applied to reach an overall filter support size of $n$. More precisely, each intermediate layer aggregates messages from the neighbors of the previous layer. In the last layer, an aggregation layer combines the features of each pixel from all the intermediate layers. 

In this paper, we follow a similar idea and investigate the best aggregation strategy for our spherical convolution. We consider 3 different strategies for the last aggregation layer:

\begin{itemize}
    \item \textbf{Concatenation} aggregation uses the feature maps of all preceding convolutional block outputs by concatenating them together \cite{xu_representation_2018}:
\begin{equation}\label{eq:concat_skip_connection}
\mathbf{z}_i^{(1)} \, \Vert \, \ldots \, \Vert \, \mathbf{z}_i^{(n)},
\end{equation}
where $\Vert$ represents the operation of concatenation.
    \item \textbf{Max} aggregation selects the most informative layer for each pixel \cite{xu_representation_2018}
\begin{equation}\label{eq:maxpool_skip_connection}
\max \left( \mathbf{z}_i^{(1)}, \ldots, \mathbf{z}_i^{(n)} \right).
\end{equation}
    \item \textbf{Addition} adopts summation to connect all preceding feature maps, as in residual networks \cite{he_deep_2016},
\begin{equation}\label{eq:sum_skip_connection}
\sum_{l=1}^n \mathbf{z}_i^{(l)}.
\end{equation}
\end{itemize}

The choice of the aggregation method may eventually depend on the target task. 
In the compression scenario investigated in the next Section, we have proven that the addition strategy is the most performing approach.

Another advantage of the recursive implementation of $n$-hop neighborhood filtering lies in the memory usage. Indeed, with HEALPix, the access to the neighbor indices of a pixel is not direct and needs to be stored. With this implementation, only the 1-hop neighborhood indices need to be stored.   

\subsection{Stride in convolution}\label{sec:stride}

In classical CNN architectures, the stride controls how the filter moves over the input signal during convolution. Stride enables to reduce the amount of computation and perform both convolution and downsampling operations simultaneously. However, defining stride on a sphere is not straightforward since a simple downsampling rule such as ``one pixel over two'' does not exist.

In our OSLO solution, we resort to the HEALPix \emph{nested} numbering scheme \cite{Gorski_2005} to define stride. This scheme arranges the pixel indices in 12 tree structures, where each tree represents one pixel of the coarse base resolution. Finer pixelization is obtained by splitting each pixel into 2 $\times$ 2 pixel structures. The embedded pixel indexing is obtained by labeling a child pixel through the concatenation of the parent index and two additional bits. For example, pixel index $5$~($101$, in binary notation) of the base HEALPix resolution (\cref{subfig:healpix_res_0}) is the parent of pixels $[20 \cdots 23]$ ($101b_1b_2$ in binary notation) in the first resolution (\cref{subfig:healpix_res_1}).
This way, the data elements that are nearby on the sphere surface are also nearby in the tree structure of the database; hence the neighbor searches are conducted efficiently in memory.

Based on the above scheme, we define a stride of $n \times n$ on the sphere by sliding the filter at every $n^2$ pixels, \emph{i.e.}, visiting pixels with a step of $n^2$. Due to the HEALPix subdivision scheme, which divides each parent into 4 children, $n$ must be a power of 2. For convolutions with hop greater than~1~ (\cref{subsec:extension_hop}), all intermediate layers~(conv$_1$ to conv$_2$ in \cref{fig:hop_extension}) are convolved with stride 1.  Finally, only the last layer (conv$_3$ in \cref{fig:hop_extension}) is convolved with stride $n \times n$, which is then combined with subsampled versions of the previous layers in the aggregation step.
We adopt such a strategy of not using stride in the intermediate layers so that the message passing algorithm can successfully share the information of all the n-hop pixels until the last convolution layer.
Examples of $2 \times 2$ strides for different resolutions are shown in \cref{fig:healpix_hierarchy_and_nesting}. Pixel indices in red indicate the visiting pixels where the convolution filter will be placed.

\subsection{Pooling/Unpooling}

\subsubsection{Pooling} 
\begin{figure} 
    \centering
    \subfloat[\label{subfig:healpix_res_0}]{%
	    \fontsize{7pt}{6pt}\selectfont
		\def\svgwidth{0.31\linewidth}
		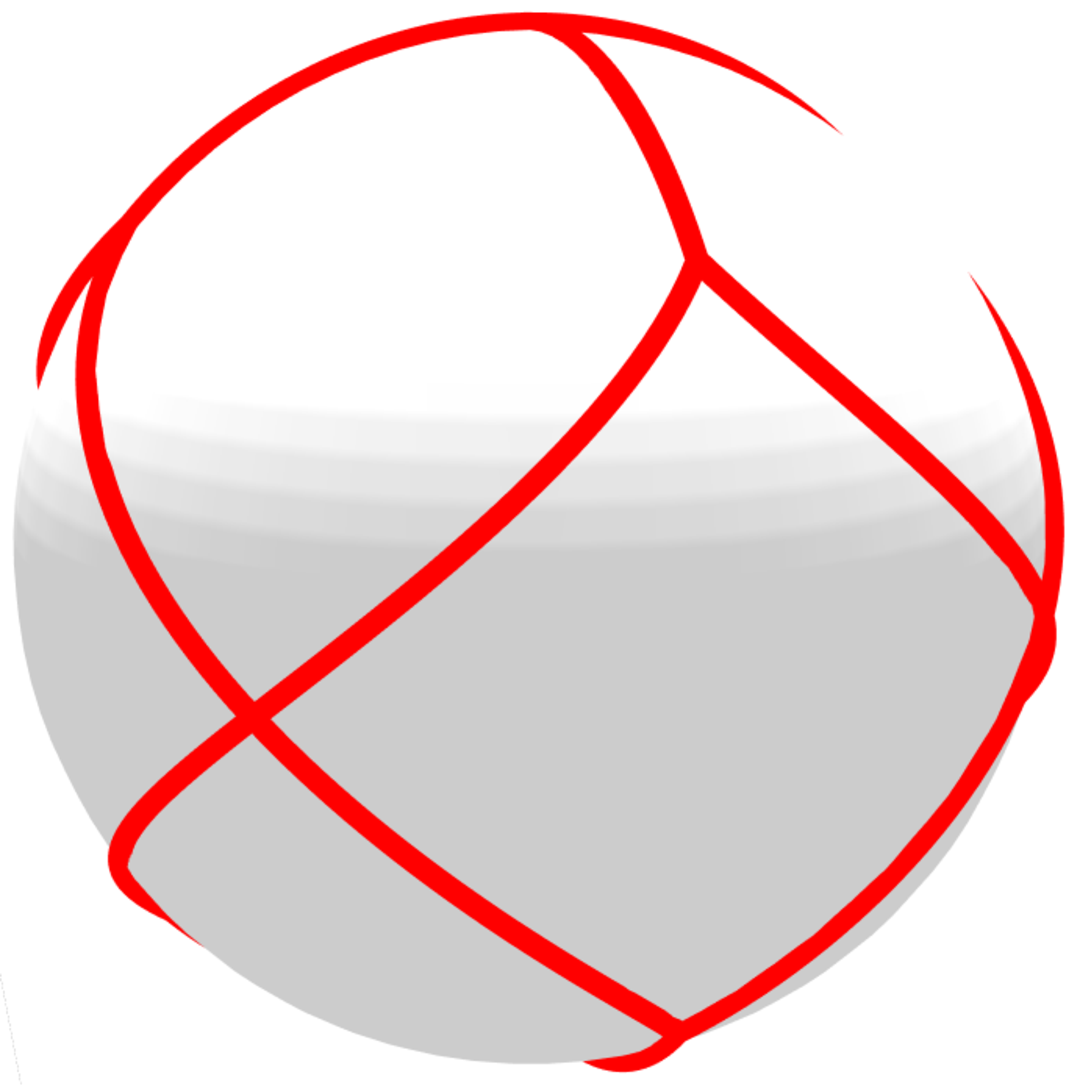}
    \hfill
    \subfloat[\label{subfig:healpix_res_1}]{%
	    \fontsize{7pt}{6pt}\selectfont
		\def\svgwidth{0.31\linewidth}
		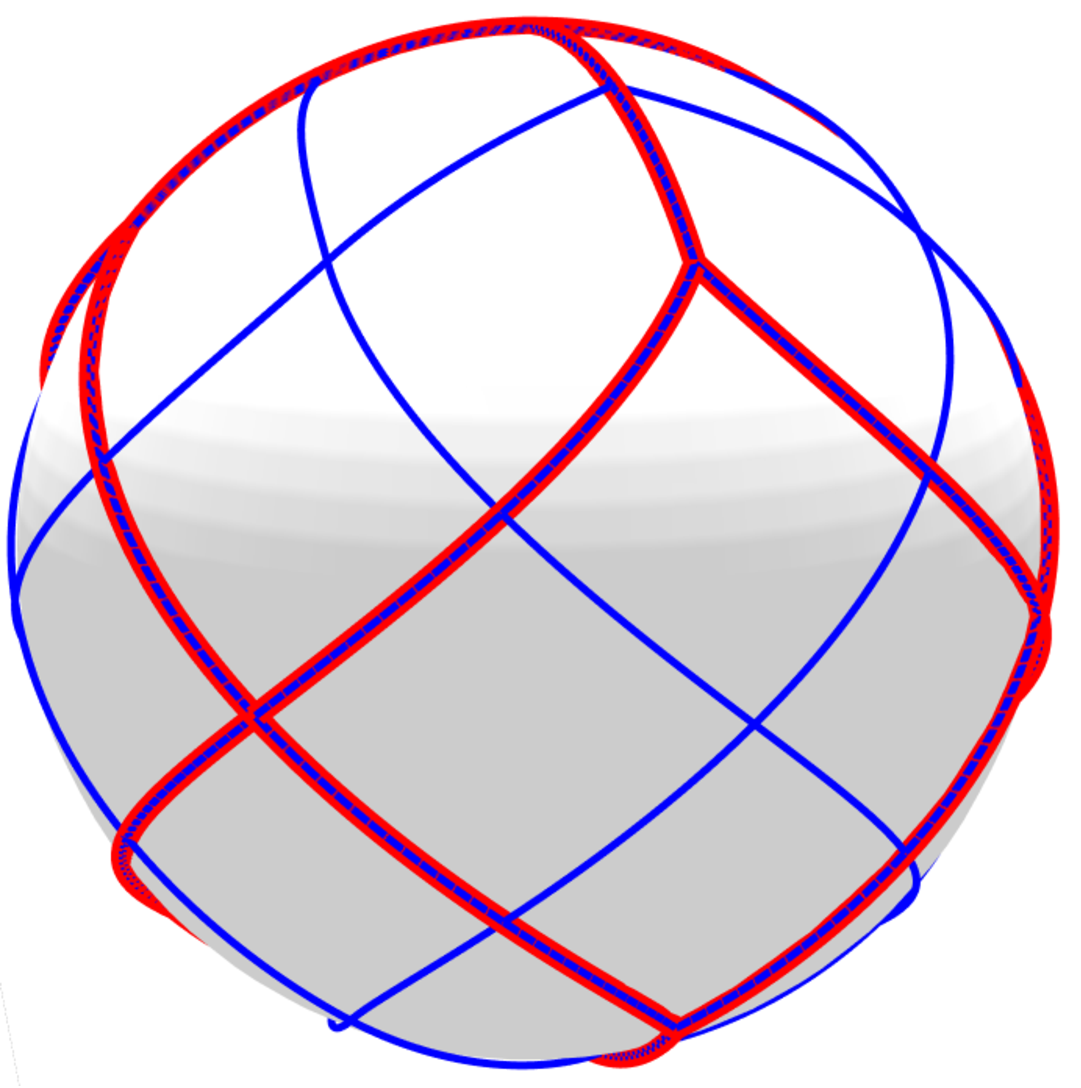}
	\hfill
    \subfloat[\label{subfig:healpix_res_2}]{%
	    \fontsize{6pt}{6pt}\selectfont
		\def\svgwidth{0.31\linewidth}
		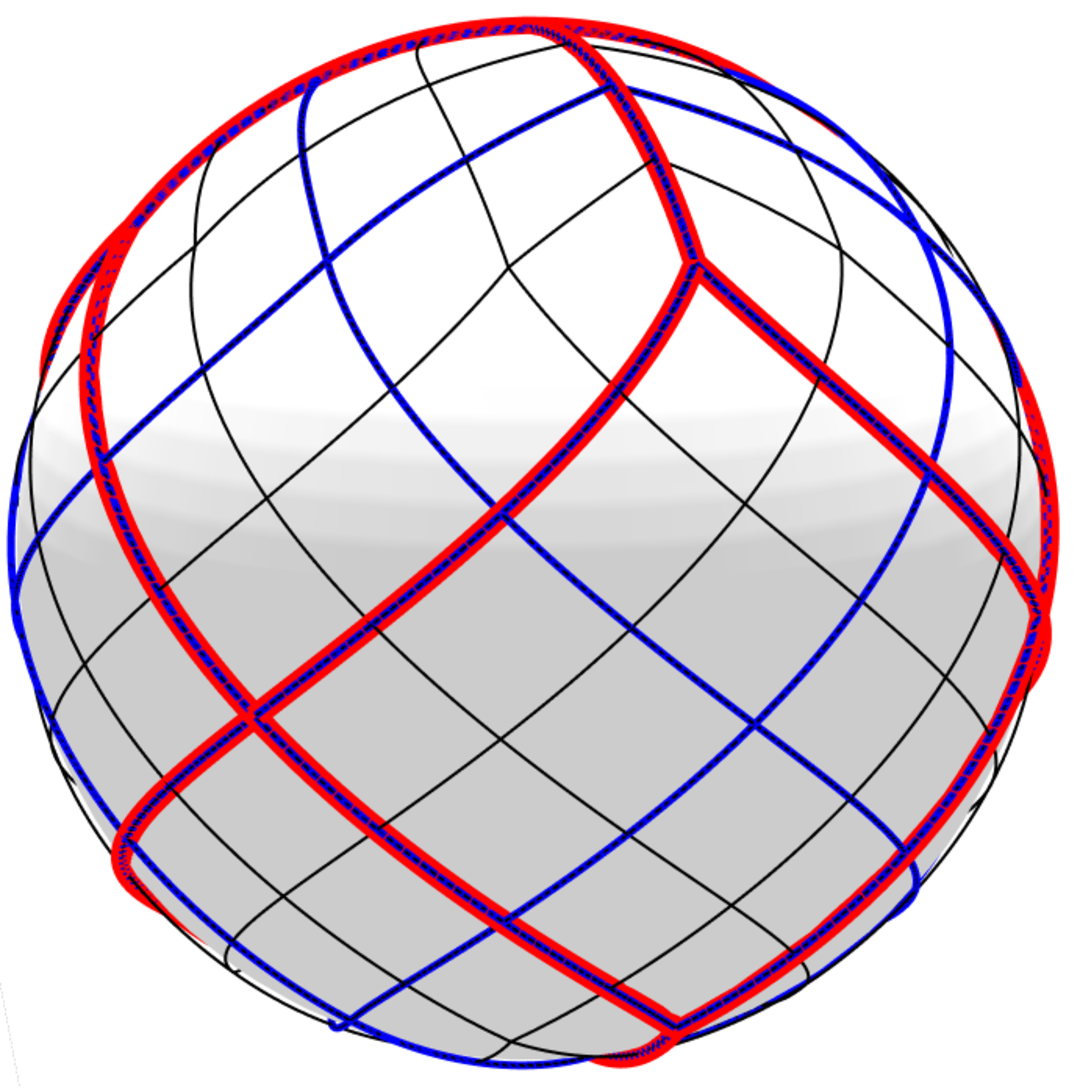}

  \caption{Hierarchical pixelization of the sphere together with the nested pixel numbering scheme. (a) HEALPix base resolution. Red lines show base-pixel boundaries. (b) HEALPix first resolution each base pixel is divided into 2 $\times$ 2 pixels. Blue lines indicate the boundaries of child pixels for base-resolution (c) HEALPix second resolution. The black lines indicate the boundaries of child pixels. The pixel indices where the center of convolution filter is placed on them to have a stride of $2 \times 2$ are shown in red.}
  \label{fig:healpix_hierarchy_and_nesting} 
\end{figure}

is a multipurpose tool in CNN. It reduces the size of the feature maps and reduces overfitting by being robust to local variations. Additionally, it tends to compute orientation and position invariant features at least locally.
Pooling can be easily implemented by exploiting the hierarchical structure of HEALPix sampling. Indeed, as shown in \cref{fig:healpix_hierarchy_and_nesting}, 
the sampling is constructed from low to high resolution, where a pixel at low resolution corresponds to four equal-area sub-pixels in the next higher resolution. Therefore, for pooling, we apply the process in reverse order and merge the child pixels into one value, assigned to the parent pixel. Formally, let $\mathcal{C}(i)$ denote the set of children of the parent pixel $i$. The pooling function $\mathbb{R}^{N_{pix} \times D} \to \mathbb{R}^{N^{\prime}_{pix} \times D}$ 
performs for each feature map, $ \forall \ d \in [1 .. D],$ and at every pixel position $\forall i \in [1 .. N^{\prime}_{pix}]$ the following operation:
\begin{equation}
y_{i,d} = f(\{x_{j,d} \mid j \in \mathcal{C}(i) \}),
\end{equation}
where $D$ is the number of feature maps, $N_{pix}$ and $N^{\prime}_{pix}$ are the numbers of pixels before and after pooling, respectively \cite{perraudin_deepsphere_2019}, and $f$ is a function that operates on a set (for instance $f$ may compute the maximum or the average over the set). Note that the resulting subsampling factor is a power of 4: $N^{\prime}_{pix}/N_{pix}=4^n, n \in \mathbb{N}$, since the HEALPix subdivision scheme divides each parent pixel into 4 sub-pixels.

\subsubsection{Unpooling~(with sub-pixel convolution)} \label{subsec:sub-pixel_conv}
is the inverse of pooling, \emph{i.e.}, a layer for increasing input resolution, and is necessary to reconstruct the input data. Unpooling can either be achieved by transpose convolution or sub-pixel convolution. Since both constructions achieve similar performance~\cite{shi2016deconvolution}, we restrict the presentation to on-the-sphere \emph{sub-pixel convolution}. However, similar ideas can be deployed to generalize transpose convolution to the sphere. 

\begin{figure}
\centering
	\fontsize{7pt}{3pt}\selectfont
	\def\svgwidth{\linewidth}
	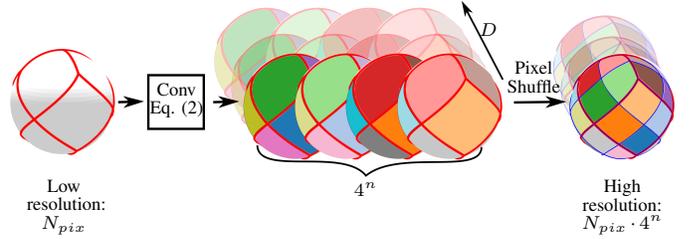
    \caption{Sub-pixel convolution on the sphere. Convolution is applied on the low-resolution sphere to produce ``$4^n \cdot D$" feature maps/channels ($n, D \in \mathbb{N}$). Then, the output feature maps are partitioned into $D$ sets of $4^n$ feature maps, and in each set, the $4^n$ channels of the same parent pixel are organized into a specific channel of child pixels.}
	\label{fig:sub-pixel_conv} 
\end{figure}

Sub-pixel convolution technique can be interpreted as a \emph{standard convolution in low-resolution space} followed by a \emph{periodic shuffling operation} known as pixel shuffle layer \cite{shi_real-time_2016}. 
Now let us consider the convolution output that is still in low-resolution and made of $4^n \times D$ channels with $N_{pix}$ number of pixels, where $D, n \in \mathbb{N}$ . A pixel shuffle transformation consists of reorganizing it to $D$ channels with $4^n \times N_{pix}$ number of pixels.
Formally, this function can be written as:
\begin{equation}\label{eq:pixel_shuffle}
\mathbb{R}^{N_{pix} \times 4^n \cdot D} \to \mathbb{R}^{N_{pix} \cdot 4^n \times D}.
\end{equation}
Defining such an unpooling operation for the spherical images is not straightforward. Our construction takes advantage of the hierarchical HEALPix sampling. We propose to first perform convolution as in~\eqref{eq:conv} to produce $L_{out}=4^n \cdot D$ feature maps with all exactly the same low resolution as the input feature maps, where $D, n \in \mathbb{N}$. Then, the $L_{out}$ output feature parts are partitioned into $D$ sets of $4^n$ maps each. Then, pixel shuffling is implemented. More precisely, in each set, the $4^n$ maps with $N_{pix}$-resolution are combined into a single map with $N'_{pix}$-resolution, where $N'_{pix}=4^n \ N_{pix}$. This is achieved by assigning to a high-resolution-block of size $4^n$, the pixel values of their low-resolution-parents in the $4^n$ maps of the set, as illustrated in \cref{fig:sub-pixel_conv}.


\subsection{Patching}\label{subsec:patch}
A patch is a subsection of a signalm \emph{e.g.}, a 2D crop of an image. In CNNs, due to GPU memory limitations and to speed up the training process, random patches sampled from higher resolution inputs are commonly used.  Due to the local properties of convolutions, patches can be successfully used to train the network, while the full resolution input can be used during inference. Since omnidirectional images are commonly much higher resolution than traditional images, patching is a must-have feature in our OSLO framework so that we can efficiently train OSLO-based networks using current GPUs.
The hierarchical property of HEALPix is used to define patches. To define a random patch of size $2^n$ by $2^n$ on the sphere sampled with HEALPix at resolution $r$, one needs to randomly choose a parent pixel on the sphere sampled at resolution $r-n$ and apply CNN to its corresponding child pixels. For instance, to define random patches of size 4 by 4 pixels for the second resolution of HEALPix (\cref{subfig:healpix_res_2}), let's assume pixel 5 is randomly selected at the base resolution (\cref{subfig:healpix_res_0}). Then, CNN is applied to all its 16 child pixels $[80 \cdots 95]$ at the second resolution. The neighboring pixels that are outside the patch do not contribute in convolution operation, \emph{i.e.}, for a pixel $i$ in the patch boundary, if its adjacent pixel $j$ is outside of the patch, $w_{j,i}$ is set to zero in Eq.~\eqref{eq:conv}.
 
Thanks to the above proposed CNN modules, OSLO supports the development of several kinds of architectures for different applications related to 360$^\circ$ images. In other words, by redefining, on-the-sphere, all the blocks commonly used in 2D CNN, many currently successful architectures applied in 2D images can be adapted to omnidirectional images as well. As an example, the next section highlights the benefits of the OSLO solution in the omnidirectional image compression application.

\section{Application to end-to-end omnidirectional image compression}
\label{sec:architecture}

\subsection{OSLO compression model}
We evaluate our novel OSLO framework in the context of end-to-end deep image compression. We use state-of-the-art representation learning frameworks that perform well for regular images, and re-define such architecture for omnidirectional images, using the on-the-sphere learning operators defined in Section~\ref{sec:spherical_conv}. For that purpose, we use the two well-known 2D deep learning image compression architectures of \cite{balle_end--end_2017,balle_variational_2018}. They are two auto-encoder-like architectures, and they enable extremely effective coding performance for perspective 2D images. Note that our methodology could have been deployed for any reference 2D compression architecture \cite{hu_learning_2021,cheng_learned_2020}.

\begin{figure} 
    \centering
    \subfloat[Architecture of \cite{balle_end--end_2017}]{
	\fontsize{8pt}{6pt}\selectfont
	\def\svgwidth{\linewidth}
	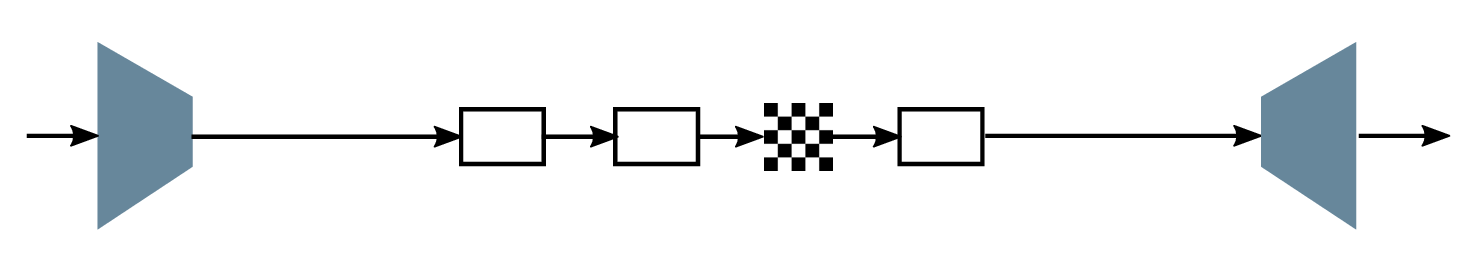
	}
	
	\subfloat[Architecture of \cite{balle_variational_2018} \label{subfig:scale_hyperprior_overview}]{
	\fontsize{8pt}{6pt}\selectfont
	\def\svgwidth{\linewidth}
	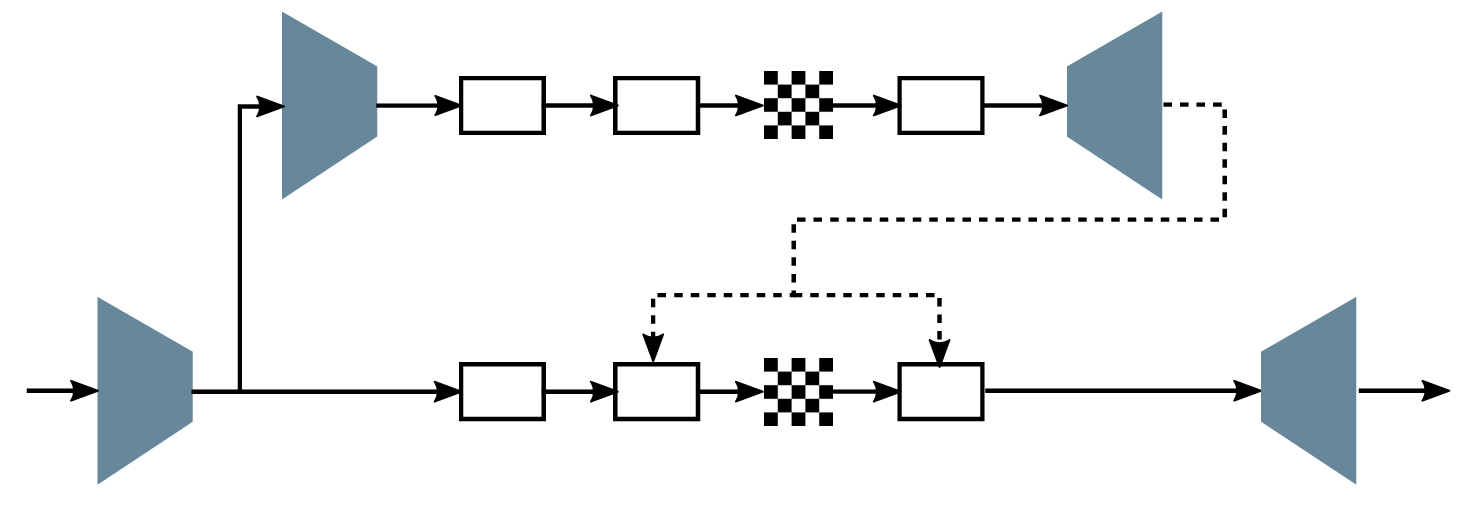
	}
  	\caption{Overview of the two network architectures used in our experiment \cite{balle_end--end_2017,balle_variational_2018}. Q represents quantization, and AE, AD represent arithmetic encoder and arithmetic decoder, respectively.}
	\label{fig:balle_overview} 
\end{figure}

The model of ~\cite{balle_variational_2018} is an extension of \cite{balle_end--end_2017} that explicitly estimates the entropy model with hyperpriors to effectively capture spatial dependencies in the latent representation. \cref{fig:balle_overview} shows an overview of the two architectures. 
They involve two types of autoencoders: (1) an image autoencoder architecture $(e,d)$ in both \cite{balle_end--end_2017} and \cite{balle_variational_2018}, (2) a hyperprior network $(e_s,d_s)$ only in \cite{balle_variational_2018}. The image encoder uses a parametric analysis transform $e$ (encoder) to transform the image vector $\mathbf{x}$ into a latent representation $\mathbf{y}$, which is subsequently quantized to $\hat{\mathbf{y}}$. In order to make the network end-to-end trainable, the quantization component, which is not differentiable by nature, is approximated by an additive uniform noise \cite{balle_end--end_2016} in the training phase. The additional hyperprior network in ~\cite{balle_variational_2018} estimates the likelihood of $\mathbf{y}$ with a Gaussian entropy model. For that, $\mathbf{y}$ is fed into $e_s$ (encoder of statistics) to summarize the distribution of standard deviations in $\pmb{\nu}$, which is then quantized into $\hat{\pmb{\nu}}$ and losslessly compressed as side information with an entropy coder. Then $\hat{\pmb{\nu}}$ is fed into $d_s$ (decoder of statistics) to estimate the spatial distribution of standard deviations $\hat{\mathbf{\sigma}}$. Finally, the image autoencoder uses $\hat{\mathbf{\sigma}}$ to losslessly compress $\hat{\mathbf{y}}$. The decoder also uses $\hat{\mathbf{\sigma}}$ to successfully recover $\hat{\mathbf{y}}$. It then feeds $\hat{\mathbf{y}}$ into a parametric synthesis transform $d$ (decoder) to obtain the reconstructed image $\hat{\mathbf{x}}$. The loss function of the model eventually maximizes the reconstruction quality and minimizes the bit rate:

\begin{equation}\label{eq:balle_loss}
\mathcal{L} = D(\mathbf{x}, \hat{\mathbf{x}}) + \lambda \cdot R(\hat{\mathbf{y}}, \hat{\pmb{\nu}}),
\end{equation}
where $D(.,.)$ indicates the distortion between the input and the reconstruction, $R$ represents the bit rate, and $\lambda$ is a scalar that balances the reconstruction quality and the bit rate. It is used to set the step of the quantization applied to the latent description.

\begin{figure*} 
    \centering
    \subfloat[Architecture of \cite{balle_end--end_2017}]{
	\fontsize{8pt}{6pt}\selectfont
	\def\svgwidth{\linewidth}
	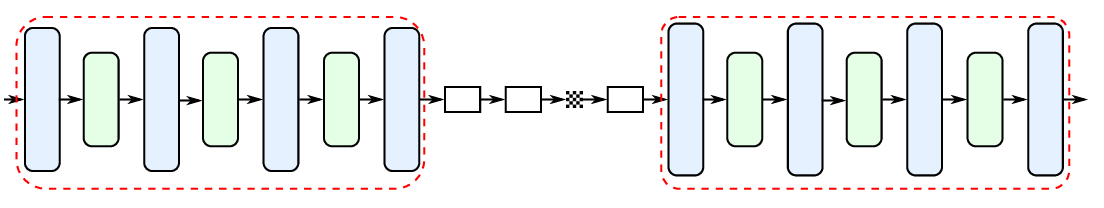
	}
	
	\subfloat[Architecture of \cite{balle_variational_2018}]{
	\fontsize{8pt}{6pt}\selectfont
	\def\svgwidth{\linewidth}
	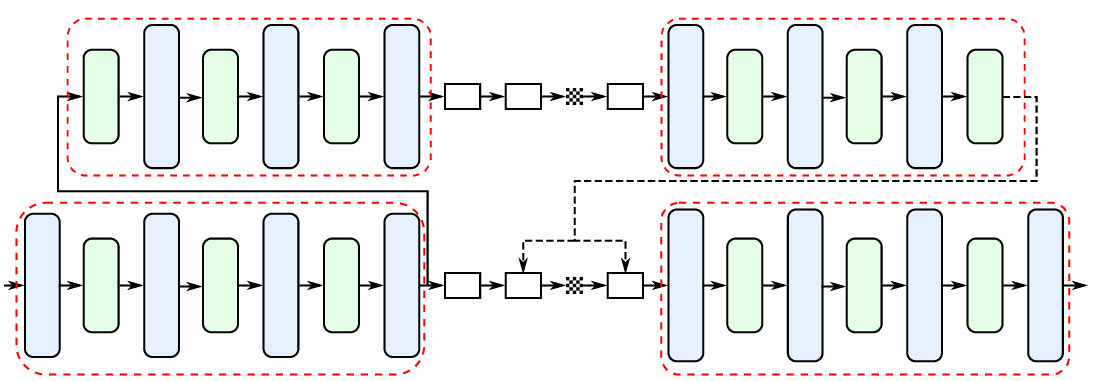
	}
  	\caption{Equivalent architectures of \cite{balle_end--end_2017,balle_variational_2018} for spherical images. Convolution parameters are denoted as: ``conv h$k$, $\downarrow l$ $\times$ N" representing a $k$-hop convolution (see \cref{fig:hop_extension}), downsampled by $l$ $\times$ number of filters. Similarly, sub-pixel convolution parameters are denoted as: ``spconv h$k$, $\uparrow l$ $\times$ N" representing a $k$-hop convolution, followed by upsampling of $l$ with pixel shuffle transformation (see \cref{fig:sub-pixel_conv}) $\times$ number of filters.}
	\label{fig:spherical_balle} 
\end{figure*}

We transform the two architectures on the sphere using the OSLO framework, as shown in \cref{fig:spherical_balle}. The 3$\times$3 and 5$\times$5 2D convolutional filters correspond to 1-hop and 2-hop filters on the sphere, respectively. For the 2-hop layer aggregation strategy, we use addition aggregation of \eqref{eq:sum_skip_connection} as it provides better performance than concatenation and max aggregations (see ablation study in \cref{subsec:aggregation_experiment}). A 2$\times$2 2D stride operator is equivalent to stride 4 on the sphere, as explained in \cref{sec:stride}. GDN/IGDN nonlinearities are transformations that are suitable for density modeling and compression of images utilizing local divisive normalization \cite{balle_density_2016}. Since the GDN/IGDN are applied pixel-wise, we use them without modifications.

\subsection{Experimental setup} \label{sec:experiments}
We evaluate the performance of on-the-sphere learning modules on the compression of 360$^\circ$ images under different conditions. We use 2170 images from the SUN360 equirectangular image database for our experiments \cite{Jianxiong_recognizing_2012}. All experiments use the same split of the dataset for learning and evaluation, consisting of 1737 images for training (80\%), 10 images for validation (0.5\%), and 423 images for testing (19.5\%). To have a fair comparison, we train all CNN models (our model and baselines) using this dataset from scratch (no pre-trained model is used).

We compare our OSLO solution with different baselines. On the one hand, we consider the solutions of \cite{balle_end--end_2017,balle_variational_2018} trained and tested on ERP. 
On the other hand, we consider the same
architecture, as in Fig.~\ref{fig:spherical_balle}, defined on HEALPix sampling, using
DeepSphere’s architecture, \textit{i.e.}, graph-based convolution and max pooling \cite{perraudin_deepsphere_2019}.
To increase the local support of the Deep sphere convolutional filters, the extension to greater than one hop is implemented by increasing the Chebyshev filter size as suggested in \cite{perraudin_deepsphere_2019,defferrard_convolutional_2016} instead of the strategy proposed for OSLO in \cref{subsec:extension_hop}. Additionally, as no upsampling strategy is introduced in DeepSphere, we use our proposed sub-pixel convolution (\cref{subsec:sub-pixel_conv}) for DeepSphere to have a fair comparison. Finally, we also consider the classical JPEG image compression algorithm on ERP for the sake of completeness. 

For the evaluation, we use the testing procedure recommended by JVET for 360$^\circ$ image compression \cite{chen_recent_2018,boyce_jvet_2017}: The high-resolution SUN360 equirectangular images of size 9104$\times$4552 are considered as the ground truth for quality evaluation. To remove the unfair bias due to the fact that the ground-truth images are available in ERP format, we downsample ERP images to a resolution of 4992$\times$2496 and HEALPix to 12582912 pixels, such that both have almost the same number of pixels (equivalent to 4K resolution), as suggested in \cite{boyce_jvet_2017,yu_a-framework_2015}. Considering that 360$^\circ$ images are spherical signals by nature, the Peak signal-to-noise ratio (PSNR) does not reflect the actual 360$^\circ$ image quality. Thus, we use \emph{Spherical PSNR}~(S-PSNR)~\cite{yu_a-framework_2015} and \emph{Weighted to Spherically uniform PSNR}~(WS-PSNR)~\cite{sun_weighted_2017} as the objective quality evaluation.

In S-PSNR the ground-truth and the decompressed images are first mapped to a sphere that is uniformly sampled with 655,362 points \cite{yu_a-framework_2015}. To ensure a fair comparison, the uniform sampling used for the S-PSNR is different from HEALPix sampling. Then the mean error between the original and decompressed signals sampled with these uniformly distributed points is calculated to compute the PSNR on the sphere. Bilinear interpolation is used to calculate values in fractional pixels. The S-PSNR value can approximate the average quality of all possible views presented to the observers. Alternately, WS-PSNR does not require remapping, and the error of each pixel is weighted to obtain the equivalent spherical area in the observation space~\cite{sun_weighted_2017}.

To train the 2D models, crops of randomly positioned patches of 256$\times$256 pixels from the training images are extracted as suggested in \cite{balle_variational_2018}. Similarly, our proposed model is also trained by randomly cropping the sphere into a region of the same size, as explained in \cref{subsec:patch}. We use the Pytorch implementation of the architecture of \cite{balle_end--end_2017,balle_variational_2018} given in \cite{begaint_compressai_2020} for training the 2D ERP models. We modify and adapt this library to implement our spherical equivalent. In all learning-based models, we use batches of 10 training images to perform stochastic gradient descent with the Adam algorithm \cite{kingma_adam_2015} and learning rate of $10^{-4}$. The training is performed for 1000 epochs, and the validation set is introduced at epoch 800 to dynamically reduce the learning rate by factor of $0.316$ when the validation loss is no longer improving more than 0.0001 for 10 epochs.

\subsection{Rate-Distortion results}

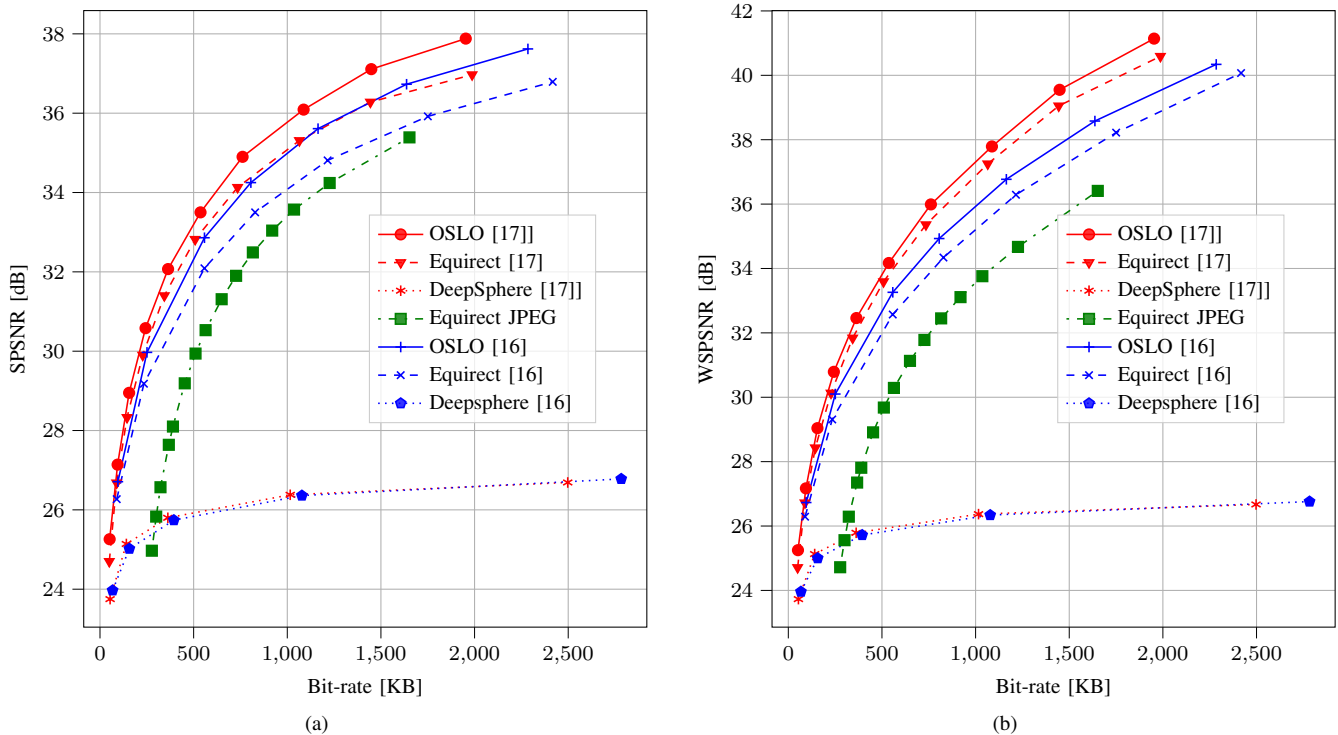
\begin{figure*} 
\centering
\subfloat[\label{subfig:SPSNR_2D_vs_HEALPIX}]{
		\centering
        \fontsize{8pt}{1pt}\selectfont
        \setlength\figurewidth{0.5\linewidth}
        \setlength\figureheight{0.4\textheight}
\begin{tikzpicture}

\definecolor{darkgray176}{RGB}{176,176,176}
\definecolor{green}{RGB}{0,128,0}
\definecolor{lightgray204}{RGB}{204,204,204}

\begin{axis}[
height=\figureheight,
legend cell align={left},
legend style={
  fill opacity=0.8,
  draw opacity=1,
  text opacity=1,
  at={(0.91,0.5)},
  anchor=east,
  draw=lightgray204
},
tick align=outside,
tick pos=left,
width=\figurewidth,
x grid style={darkgray176},
xlabel={Bit-rate [KB]},
xmajorgrids,
xmin=-86.9228411865234, xmax=2919.77401550293,
xtick style={color=black},
y grid style={darkgray176},
ylabel={SPSNR [dB]},
ymajorgrids,
ymin=23.0435, ymax=38.5865,
ytick style={color=black}
]
\addplot [semithick, red, mark=*, mark size=2, mark options={solid}]
table {%
51.5357287597656 25.26
94.0177026367188 27.14
155.46451171875 28.95
243.177618408203 30.58
363.517508544922 32.07
536.791380615234 33.5
761.332622070312 34.9
1087.21661621094 36.09
1448.59866333008 37.11
1952.96680175781 37.88
};
\addlegendentry{OSLO [17]]}
\addplot [semithick, red, dashed, mark=triangle*, mark size=2, mark options={solid,rotate=180}]
table {%
49.7451977539062 24.7
88.9248852539063 26.68
144.388436279297 28.33
228.041002197266 29.9
343.580277099609 31.4
508.268247070312 32.82
734.825557861328 34.13
1064.84652099609 35.31
1443.90678588867 36.28
1986.89565795898 36.97
};
\addlegendentry{Equirect [17]}
\addplot [semithick, red, dotted, mark=asterisk, mark size=2, mark options={solid}]
table {%
54.006953125 23.75
140.795046386719 25.14
362.294437255859 25.8
1015.69516479492 26.38
2497.41734985352 26.69
};
\addlegendentry{DeepSphere [17]]}
\addplot [semithick, green, dash pattern=on 1pt off 3pt on 3pt off 3pt, mark=square*, mark size=2, mark options={solid}]
table {%
276.910405273438 24.97
300.086496582031 25.83
322.996689453125 26.57
367.487445068359 27.64
389.514713134766 28.1
452.564852294922 29.19
510.123060302734 29.94
563.884685058594 30.53
649.673094482422 31.31
726.727349853516 31.9
816.160765380859 32.49
919.450705566406 33.04
1036.05268188477 33.57
1226.33556762695 34.24
1652.37840087891 35.39
};
\addlegendentry{Equirect JPEG}
\addplot [semithick, blue, mark=+, mark size=2, mark options={solid}]
table {%
95.6473022460937 26.71
250.883117675781 29.97
557.819906005859 32.86
805.575576171875 34.25
1164.23697875977 35.61
1636.97487304688 36.73
2284.86313354492 37.62
};
\addlegendentry{OSLO [16]}
\addplot [semithick, blue, dashed, mark=x, mark size=2, mark options={solid}]
table {%
89.3442492675781 26.27
234.516705322266 29.18
557.610926513672 32.09
828.33169921875 33.5
1215.43615234375 34.81
1750.42578125 35.92
2417.47916625977 36.79
};
\addlegendentry{Equirect [16]}
\addplot [semithick, blue, dotted, mark=pentagon*, mark size=2, mark options={solid}]
table {%
67.3131921386719 23.97
156.795018310547 25.02
394.429309082031 25.74
1077.95155517578 26.36
2783.1059765625 26.78
};
\addlegendentry{Deepsphere [16]}
\end{axis}

\end{tikzpicture}}
\hfill
\subfloat[\label{subfig:WSPSNR_2D_vs_HEALPIX}]{
		\centering
        \fontsize{8pt}{1pt}\selectfont
        \setlength\figurewidth{0.5\linewidth}
        \setlength\figureheight{0.4\textheight}
\begin{tikzpicture}

\definecolor{darkgray176}{RGB}{176,176,176}
\definecolor{green}{RGB}{0,128,0}
\definecolor{lightgray204}{RGB}{204,204,204}

\begin{axis}[
height=\figureheight,
legend cell align={left},
legend style={
  fill opacity=0.8,
  draw opacity=1,
  text opacity=1,
  at={(0.91,0.5)},
  anchor=east,
  draw=lightgray204
},
tick align=outside,
tick pos=left,
width=\figurewidth,
x grid style={darkgray176},
xlabel={Bit-rate [KB]},
xmajorgrids,
xmin=-86.9228411865234, xmax=2919.77401550293,
xtick style={color=black},
y grid style={darkgray176},
ylabel={WSPSNR [dB]},
ymajorgrids,
ymin=22.8595, ymax=42.0105,
ytick style={color=black}
]
\addplot [semithick, red, mark=*, mark size=2, mark options={solid}]
table {%
51.5357287597656 25.25
94.0177026367188 27.17
155.46451171875 29.04
243.177618408203 30.79
363.517508544922 32.46
536.791380615234 34.17
761.332622070312 35.99
1087.21661621094 37.79
1448.59866333008 39.55
1952.96680175781 41.14
};
\addlegendentry{OSLO [17]]}
\addplot [semithick, red, dashed, mark=triangle*, mark size=2, mark options={solid,rotate=180}]
table {%
49.7451977539062 24.71
88.9248852539063 26.72
144.388436279297 28.43
228.041002197266 30.13
343.580277099609 31.84
508.268247070312 33.59
734.825557861328 35.36
1064.84652099609 37.25
1443.90678588867 39.05
1986.89565795898 40.59
};
\addlegendentry{Equirect [17]}
\addplot [semithick, red, dotted, mark=asterisk, mark size=2, mark options={solid}]
table {%
54.006953125 23.73
140.795046386719 25.13
362.294437255859 25.79
1015.69516479492 26.37
2497.41734985352 26.67
};
\addlegendentry{DeepSphere [17]]}
\addplot [semithick, green, dash pattern=on 1pt off 3pt on 3pt off 3pt, mark=square*, mark size=2, mark options={solid}]
table {%
276.910405273438 24.72
300.086496582031 25.56
322.996689453125 26.29
367.487445068359 27.35
389.514713134766 27.81
452.564852294922 28.91
510.123060302734 29.68
563.884685058594 30.29
649.673094482422 31.13
726.727349853516 31.78
816.160765380859 32.45
919.450705566406 33.11
1036.05268188477 33.76
1226.33556762695 34.67
1652.37840087891 36.41
};
\addlegendentry{Equirect JPEG}
\addplot [semithick, blue, mark=+, mark size=2, mark options={solid}]
table {%
95.6473022460937 26.73
250.883117675781 30.1
557.819906005859 33.26
805.575576171875 34.93
1164.23697875977 36.77
1636.97487304688 38.58
2284.86313354492 40.34
};
\addlegendentry{OSLO [16]}
\addplot [semithick, blue, dashed, mark=x, mark size=2, mark options={solid}]
table {%
89.3442492675781 26.29
234.516705322266 29.31
557.610926513672 32.57
828.33169921875 34.34
1215.43615234375 36.29
1750.42578125 38.22
2417.47916625977 40.07
};
\addlegendentry{Equirect [16]}
\addplot [semithick, blue, dotted, mark=pentagon*, mark size=2, mark options={solid}]
table {%
67.3131921386719 23.96
156.795018310547 25
394.429309082031 25.72
1077.95155517578 26.34
2783.1059765625 26.76
};
\addlegendentry{Deepsphere [16]}
\end{axis}

\end{tikzpicture}}
\caption{Rate-distortion curves aggregated over the test images. The solid, dashed, dotted, and dash-dotted lines represent OSLO implementation, 2D deep learning architectures on Equirectangular projections, Deepsphere architectures, and Equirectangular JPEG compression. Red color refers to the architecture of \cite{balle_variational_2018}, blue color indicates \cite{balle_end--end_2017} architecture, and green color indicates JPEG compression. (a) S-PSNR. (b) WS-PSNR.}
\label{fig:2D_vs_HEALPIX} 
\end{figure*}

Given a test image set denoted by $\mathcal{X}$, we compute for each $i\in \mathcal{X}$ and for several $\lambda$, the rate-distortion pair $(R(i,\lambda),D(i,\lambda))$ achieved by each architecture of interest. The aggregated Rate-Distortion performance of each architecture is then computed as in \cite{balle_variational_2018} by averaging over $\mathcal{X}$:
\begin{align}
    R_\lambda  = \frac{1}{|\mathcal{X}|} \sum_{i \in \mathcal{X}} R(i,\lambda) \\
    D_\lambda = \frac{1}{|\mathcal{X}|} \sum_{i \in \mathcal{X}} D(i,\lambda).
\end{align}
\cref{fig:2D_vs_HEALPIX} present, for each architecture, the $(R_\lambda, D_\lambda)$ for different values of $\lambda$.
%
%
We can see that our OSLO-based approach significantly outperforms the baseline methods. The DeepSphere solution can not reconstruct the image properly, and even when the bitrate is increased, the performance gets saturated because the isotropic filters are not expressive enough to reconstruct the image. Comparing 2D learning architectures applied to ERP with JPEG compression reveals that, similar to the results given in \cite{balle_end--end_2017,balle_variational_2018} for perspective images, these models do not lose performance when the input train/test data is ERP, and they still perform much better than JPEG. We can conclude that
2D learning approaches can keep their efficiency compared to JPEG even with ERP format. However, our proposed solution is able to yet outperform the ERP representation due to the efficient learning tools of OSLO that are directly defined on the sphere.

\begin{figure*} 
\centering
\subfloat[1980.89 KB]{
		\centering
        \includegraphics[width=0.8\linewidth]{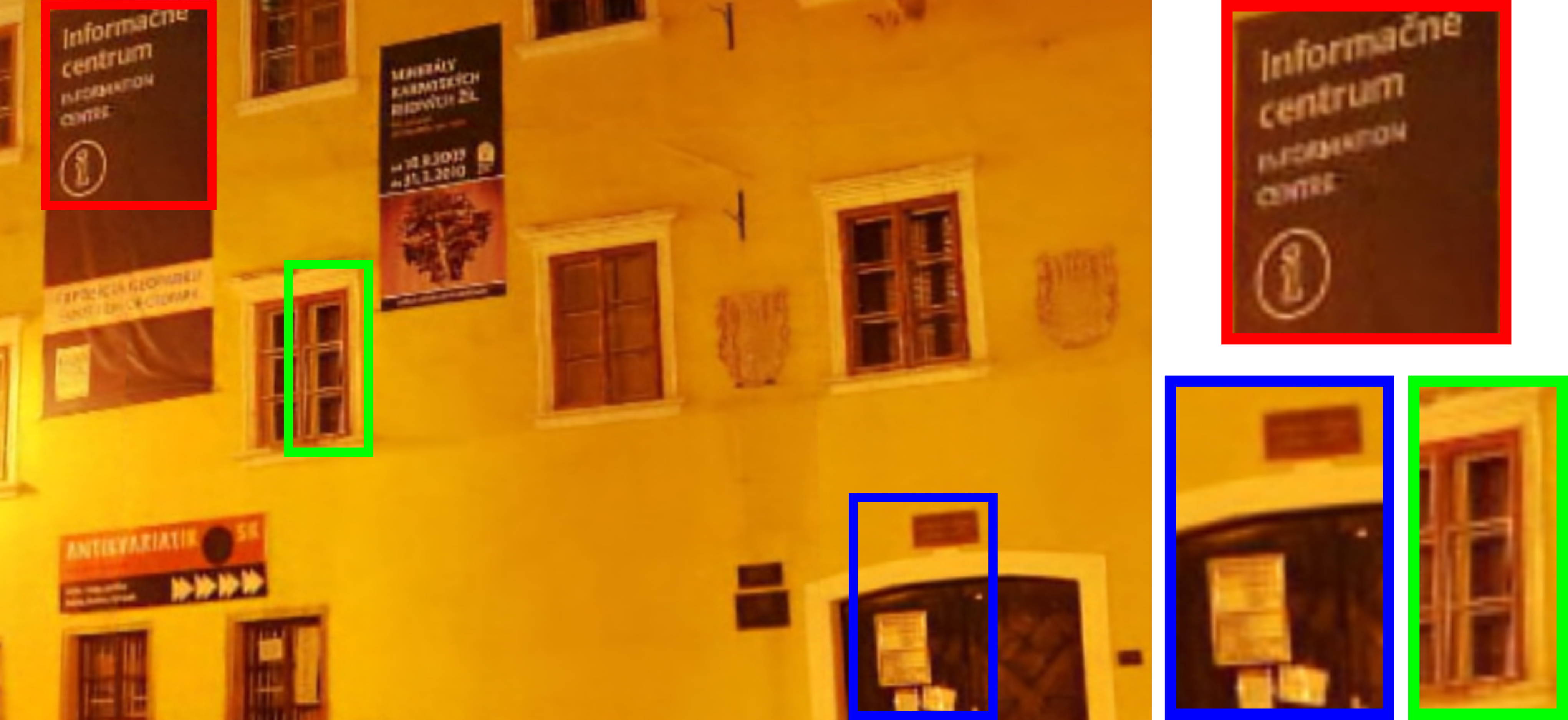}}

\subfloat[1920.86 KB]{
		\centering
        \includegraphics[width=0.8\linewidth]{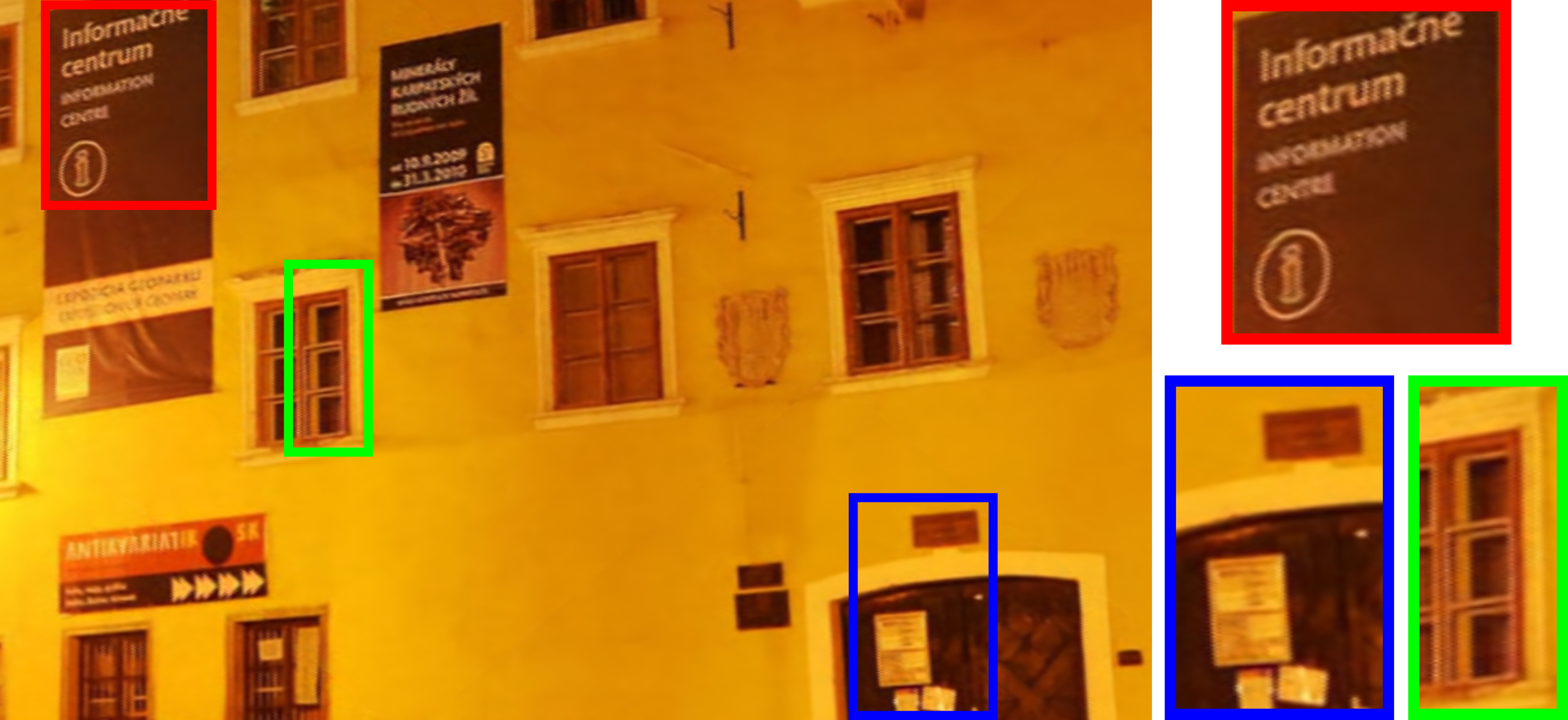}}
        
\caption{A viewport at a specific direction of spherical images compressed almost at the same rate. A zoom version of each colored rectangular region is shown on the right. (a) Equirectangular image compressed with \cite{balle_end--end_2017} in 1980.89 KB. (b) The same image compressed with OSLO in 1920.86 KB.}
\label{fig:viewport_comparison_01} 
\end{figure*}

\begin{figure*} 
\centering

\subfloat[1766.94 KB]{
		\centering
        \includegraphics[width=0.8\linewidth]{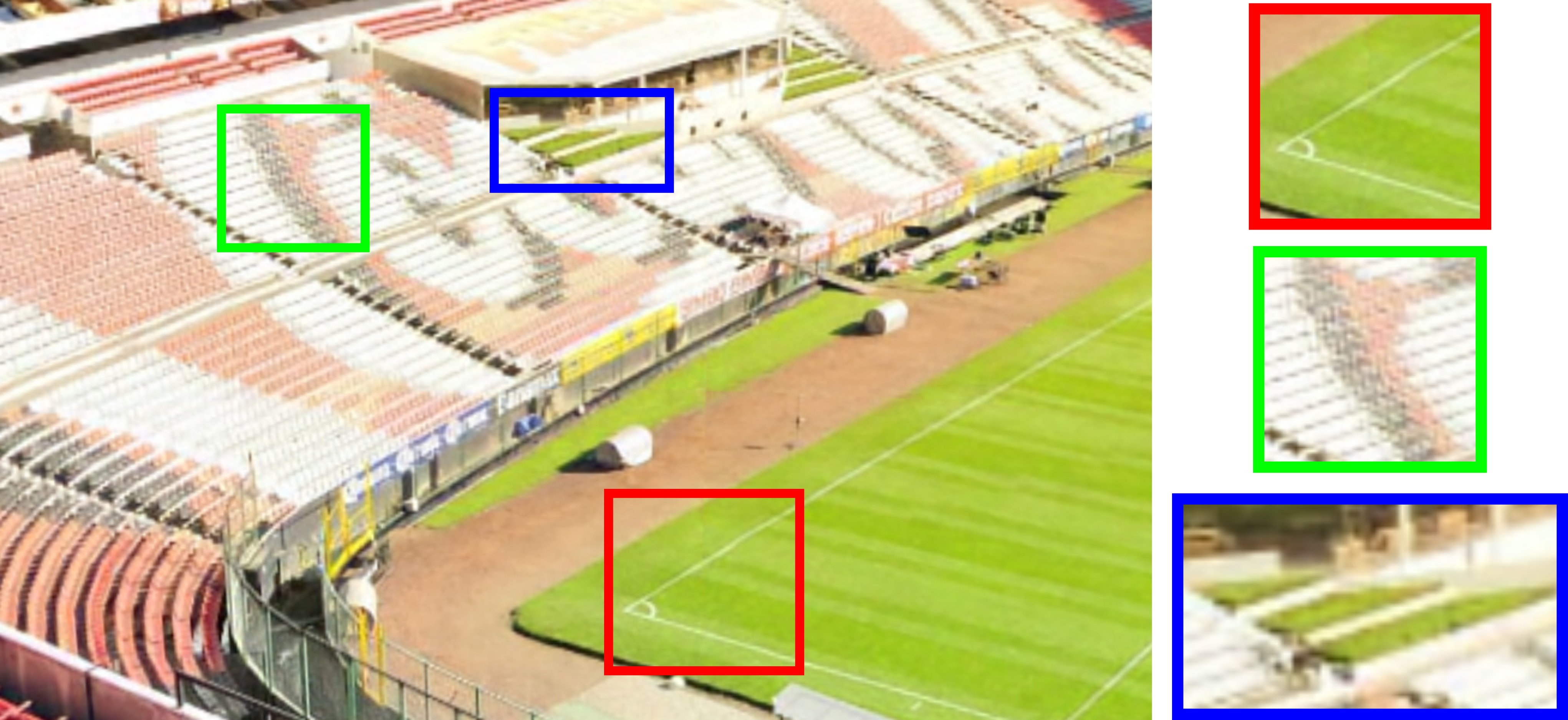}}
        
\subfloat[1613.62 KB]{
		\centering
        \includegraphics[width=0.8\linewidth]{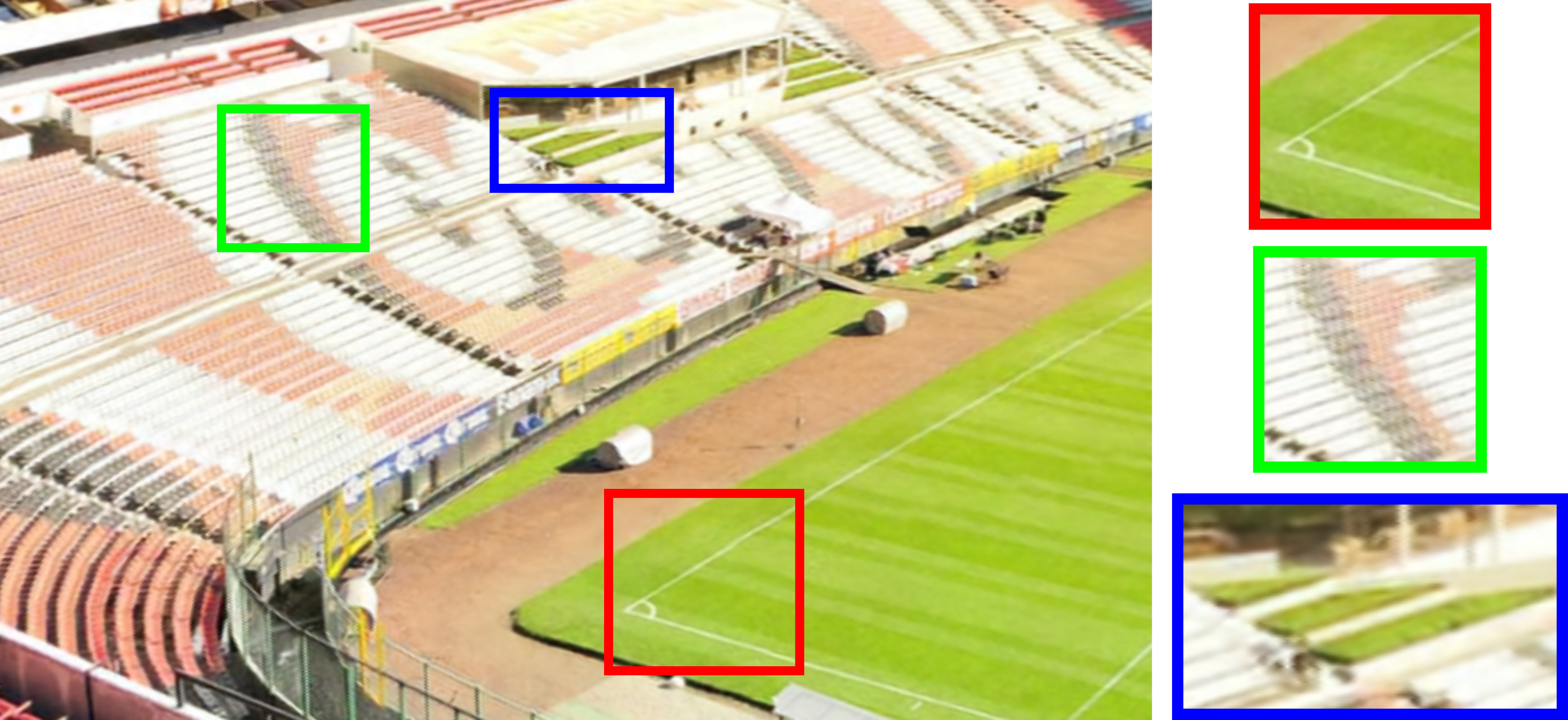}}
        
\caption{A viewport at a specific direction of spherical images compressed almost at the same rate. A zoom version of each colored rectangular region is shown on the right. (a) Equirectangular image compressed with \cite{balle_end--end_2017} in 1766.94 KB. (b) The same image compressed with OSLO in 1613.62 KB.}
\label{fig:viewport_comparison_02} 
\end{figure*}

The compressed results between OSLO and 2D deep learning architecture applied to equirectangular projection are also visually compared. Comparison between OSLO and the DeepSphere’s architecture \cite{perraudin_deepsphere_2019}, that use the same HEALPix sampling, will be studied in \cref{sec:filter_visualization}. \cref{fig:viewport_comparison_01,fig:viewport_comparison_02} illustrate viewports of spherical images (the portion of spherical images seen by HMD devices at a specific direction) that are compressed almost at the same rate with the same architecture as \cite{balle_end--end_2017}. It can be seen that an image compressed with OSLO has less distortion and keeps the detail, for example, the texts are more visible.   

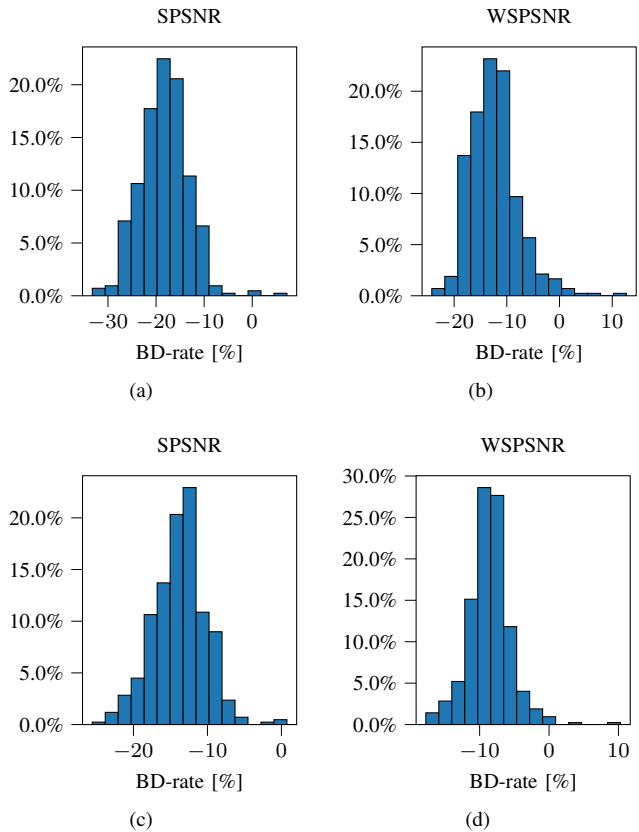
\begin{figure} 
\centering
\subfloat[]{
		\centering
        \fontsize{8pt}{1pt}\selectfont
        \setlength\figurewidth{0.5\linewidth}
        \setlength\figureheight{0.2\textheight}
\begin{tikzpicture}

\definecolor{darkgray176}{RGB}{176,176,176}
\definecolor{steelblue31119180}{RGB}{31,119,180}

\begin{axis}[
height=\figureheight,
tick align=outside,
tick pos=left,
title={SPSNR},
width=\figurewidth,
x grid style={darkgray176},
xlabel={BD-rate [\%]},
xmin=-35.2914144016599, xmax=9.24377298306577,
xtick style={color=black},
y grid style={darkgray176},
ymin=0, ymax=0.23581560283688,
ytick style={color=black},
ytick={0,0.05,0.1,0.15,0.2,0.25},
yticklabels={0.0\%,5.0\%,10.0\%,15.0\%,20.0\%,25.0\%}
]
\draw[draw=black,fill=steelblue31119180] (axis cs:-33.2670877023542,0) rectangle (axis cs:-30.5679854366132,0.00709219858156028);
\draw[draw=black,fill=steelblue31119180] (axis cs:-30.5679854366132,0) rectangle (axis cs:-27.8688831708723,0.00945626477541371);
\draw[draw=black,fill=steelblue31119180] (axis cs:-27.8688831708723,0) rectangle (axis cs:-25.1697809051313,0.0709219858156029);
\draw[draw=black,fill=steelblue31119180] (axis cs:-25.1697809051313,0) rectangle (axis cs:-22.4706786393904,0.106382978723404);
\draw[draw=black,fill=steelblue31119180] (axis cs:-22.4706786393904,0) rectangle (axis cs:-19.7715763736494,0.177304964539007);
\draw[draw=black,fill=steelblue31119180] (axis cs:-19.7715763736494,0) rectangle (axis cs:-17.0724741079085,0.224586288416076);
\draw[draw=black,fill=steelblue31119180] (axis cs:-17.0724741079085,0) rectangle (axis cs:-14.3733718421675,0.205673758865248);
\draw[draw=black,fill=steelblue31119180] (axis cs:-14.3733718421675,0) rectangle (axis cs:-11.6742695764266,0.113475177304965);
\draw[draw=black,fill=steelblue31119180] (axis cs:-11.6742695764266,0) rectangle (axis cs:-8.97516731068564,0.066193853427896);
\draw[draw=black,fill=steelblue31119180] (axis cs:-8.97516731068564,0) rectangle (axis cs:-6.27606504494469,0.00945626477541371);
\draw[draw=black,fill=steelblue31119180] (axis cs:-6.27606504494469,0) rectangle (axis cs:-3.57696277920374,0.00236406619385343);
\draw[draw=black,fill=steelblue31119180] (axis cs:-3.57696277920374,0) rectangle (axis cs:-0.877860513462792,0);
\draw[draw=black,fill=steelblue31119180] (axis cs:-0.877860513462792,0) rectangle (axis cs:1.82124175227815,0.00472813238770686);
\draw[draw=black,fill=steelblue31119180] (axis cs:1.82124175227815,0) rectangle (axis cs:4.5203440180191,0);
\draw[draw=black,fill=steelblue31119180] (axis cs:4.5203440180191,0) rectangle (axis cs:7.21944628376006,0.00236406619385343);
\end{axis}

\end{tikzpicture}}
\hfill
\subfloat[]{
		\centering
        \fontsize{8pt}{1pt}\selectfont
        \setlength\figurewidth{0.5\linewidth}
        \setlength\figureheight{0.2\textheight}
\begin{tikzpicture}

\definecolor{darkgray176}{RGB}{176,176,176}
\definecolor{steelblue31119180}{RGB}{31,119,180}

\begin{axis}[
height=\figureheight,
tick align=outside,
tick pos=left,
title={WSPSNR},
width=\figurewidth,
x grid style={darkgray176},
xlabel={BD-rate [\%]},
xmin=-26.0857085806815, xmax=14.551907917732,
xtick style={color=black},
y grid style={darkgray176},
ymin=0, ymax=0.243262411347518,
ytick style={color=black},
ytick={0,0.05,0.1,0.15,0.2,0.25},
yticklabels={0.0\%,5.0\%,10.0\%,15.0\%,20.0\%,25.0\%}
]
\draw[draw=black,fill=steelblue31119180] (axis cs:-24.2385441943899,0) rectangle (axis cs:-21.7756583460013,0.00709219858156028);
\draw[draw=black,fill=steelblue31119180] (axis cs:-21.7756583460013,0) rectangle (axis cs:-19.3127724976126,0.0189125295508274);
\draw[draw=black,fill=steelblue31119180] (axis cs:-19.3127724976126,0) rectangle (axis cs:-16.8498866492239,0.137115839243499);
\draw[draw=black,fill=steelblue31119180] (axis cs:-16.8498866492239,0) rectangle (axis cs:-14.3870008008352,0.179669030732861);
\draw[draw=black,fill=steelblue31119180] (axis cs:-14.3870008008352,0) rectangle (axis cs:-11.9241149524465,0.231678486997636);
\draw[draw=black,fill=steelblue31119180] (axis cs:-11.9241149524465,0) rectangle (axis cs:-9.46122910405779,0.219858156028369);
\draw[draw=black,fill=steelblue31119180] (axis cs:-9.46122910405779,0) rectangle (axis cs:-6.9983432556691,0.0969267139479906);
\draw[draw=black,fill=steelblue31119180] (axis cs:-6.9983432556691,0) rectangle (axis cs:-4.5354574072804,0.0567375886524823);
\draw[draw=black,fill=steelblue31119180] (axis cs:-4.5354574072804,0) rectangle (axis cs:-2.07257155889171,0.0212765957446809);
\draw[draw=black,fill=steelblue31119180] (axis cs:-2.07257155889171,0) rectangle (axis cs:0.390314289496981,0.016548463356974);
\draw[draw=black,fill=steelblue31119180] (axis cs:0.390314289496981,0) rectangle (axis cs:2.85320013788567,0.00709219858156028);
\draw[draw=black,fill=steelblue31119180] (axis cs:2.85320013788567,0) rectangle (axis cs:5.31608598627437,0.00236406619385343);
\draw[draw=black,fill=steelblue31119180] (axis cs:5.31608598627437,0) rectangle (axis cs:7.77897183466306,0.00236406619385343);
\draw[draw=black,fill=steelblue31119180] (axis cs:7.77897183466306,0) rectangle (axis cs:10.2418576830517,0);
\draw[draw=black,fill=steelblue31119180] (axis cs:10.2418576830517,0) rectangle (axis cs:12.7047435314404,0.00236406619385343);
\end{axis}

\end{tikzpicture}}
        
\subfloat[]{
		\centering
        \fontsize{8pt}{1pt}\selectfont
        \setlength\figurewidth{0.5\linewidth}
        \setlength\figureheight{0.2\textheight}
\begin{tikzpicture}

\definecolor{darkgray176}{RGB}{176,176,176}
\definecolor{steelblue31119180}{RGB}{31,119,180}

\begin{axis}[
height=\figureheight,
tick align=outside,
tick pos=left,
title={SPSNR},
width=\figurewidth,
x grid style={darkgray176},
xlabel={BD-rate [\%]},
xmin=-26.8804414648285, xmax=2.06054170418088,
xtick style={color=black},
y grid style={darkgray176},
ymin=0, ymax=0.240780141843972,
ytick style={color=black},
ytick={0,0.05,0.1,0.15,0.2,0.25},
yticklabels={0.0\%,5.0\%,10.0\%,15.0\%,20.0\%,25.0\%}
]
\draw[draw=black,fill=steelblue31119180] (axis cs:-25.5649422298735,0) rectangle (axis cs:-23.8109432499335,0.00236406619385343);
\draw[draw=black,fill=steelblue31119180] (axis cs:-23.8109432499336,0) rectangle (axis cs:-22.0569442699936,0.0118203309692671);
\draw[draw=black,fill=steelblue31119180] (axis cs:-22.0569442699936,0) rectangle (axis cs:-20.3029452900536,0.0283687943262411);
\draw[draw=black,fill=steelblue31119180] (axis cs:-20.3029452900536,0) rectangle (axis cs:-18.5489463101137,0.0449172576832151);
\draw[draw=black,fill=steelblue31119180] (axis cs:-18.5489463101137,0) rectangle (axis cs:-16.7949473301737,0.106382978723404);
\draw[draw=black,fill=steelblue31119180] (axis cs:-16.7949473301737,0) rectangle (axis cs:-15.0409483502337,0.137115839243499);
\draw[draw=black,fill=steelblue31119180] (axis cs:-15.0409483502337,0) rectangle (axis cs:-13.2869493702938,0.203309692671395);
\draw[draw=black,fill=steelblue31119180] (axis cs:-13.2869493702938,0) rectangle (axis cs:-11.5329503903538,0.229314420803783);
\draw[draw=black,fill=steelblue31119180] (axis cs:-11.5329503903538,0) rectangle (axis cs:-9.77895141041386,0.108747044917258);
\draw[draw=black,fill=steelblue31119180] (axis cs:-9.77895141041386,0) rectangle (axis cs:-8.0249524304739,0.0898345153664303);
\draw[draw=black,fill=steelblue31119180] (axis cs:-8.0249524304739,0) rectangle (axis cs:-6.27095345053394,0.0236406619385343);
\draw[draw=black,fill=steelblue31119180] (axis cs:-6.27095345053394,0) rectangle (axis cs:-4.51695447059398,0.00709219858156028);
\draw[draw=black,fill=steelblue31119180] (axis cs:-4.51695447059398,0) rectangle (axis cs:-2.76295549065402,0);
\draw[draw=black,fill=steelblue31119180] (axis cs:-2.76295549065402,0) rectangle (axis cs:-1.00895651071405,0.00236406619385343);
\draw[draw=black,fill=steelblue31119180] (axis cs:-1.00895651071405,0) rectangle (axis cs:0.74504246922591,0.00472813238770686);
\end{axis}

\end{tikzpicture}}
\hfill
\subfloat[]{
		\centering
        \fontsize{8pt}{1pt}\selectfont
        \setlength\figurewidth{0.5\linewidth}
        \setlength\figureheight{0.2\textheight}
\begin{tikzpicture}

\definecolor{darkgray176}{RGB}{176,176,176}
\definecolor{steelblue31119180}{RGB}{31,119,180}

\begin{axis}[
height=\figureheight,
tick align=outside,
tick pos=left,
title={WSPSNR},
width=\figurewidth,
x grid style={darkgray176},
xlabel={BD-rate [\%]},
xmin=-19.1525413535403, xmax=11.698715519798,
xtick style={color=black},
y grid style={darkgray176},
ymin=0, ymax=0.300354609929078,
ytick style={color=black},
ytick={0,0.05,0.1,0.15,0.2,0.25,0.3,0.35},
yticklabels={0.0\%,5.0\%,10.0\%,15.0\%,20.0\%,25.0\%,30.0\%,35.0\%}
]
\draw[draw=black,fill=steelblue31119180] (axis cs:-17.7502114956613,0) rectangle (axis cs:-15.8804383518226,0.0141843971631206);
\draw[draw=black,fill=steelblue31119180] (axis cs:-15.8804383518226,0) rectangle (axis cs:-14.0106652079839,0.0283687943262411);
\draw[draw=black,fill=steelblue31119180] (axis cs:-14.0106652079839,0) rectangle (axis cs:-12.1408920641452,0.0520094562647754);
\draw[draw=black,fill=steelblue31119180] (axis cs:-12.1408920641452,0) rectangle (axis cs:-10.2711189203066,0.15130023640662);
\draw[draw=black,fill=steelblue31119180] (axis cs:-10.2711189203066,0) rectangle (axis cs:-8.40134577646787,0.286052009456265);
\draw[draw=black,fill=steelblue31119180] (axis cs:-8.40134577646787,0) rectangle (axis cs:-6.53157263262919,0.276595744680851);
\draw[draw=black,fill=steelblue31119180] (axis cs:-6.53157263262919,0) rectangle (axis cs:-4.66179948879051,0.118203309692672);
\draw[draw=black,fill=steelblue31119180] (axis cs:-4.66179948879051,0) rectangle (axis cs:-2.79202634495183,0.0401891252955083);
\draw[draw=black,fill=steelblue31119180] (axis cs:-2.79202634495183,0) rectangle (axis cs:-0.922253201113143,0.0189125295508274);
\draw[draw=black,fill=steelblue31119180] (axis cs:-0.922253201113143,0) rectangle (axis cs:0.947519942725538,0.00945626477541371);
\draw[draw=black,fill=steelblue31119180] (axis cs:0.947519942725538,0) rectangle (axis cs:2.81729308656422,0);
\draw[draw=black,fill=steelblue31119180] (axis cs:2.81729308656422,0) rectangle (axis cs:4.68706623040291,0.00236406619385343);
\draw[draw=black,fill=steelblue31119180] (axis cs:4.68706623040291,0) rectangle (axis cs:6.55683937424159,0);
\draw[draw=black,fill=steelblue31119180] (axis cs:6.55683937424159,0) rectangle (axis cs:8.42661251808027,0);
\draw[draw=black,fill=steelblue31119180] (axis cs:8.42661251808027,0) rectangle (axis cs:10.296385661919,0.00236406619385343);
\end{axis}

\end{tikzpicture}}
                
\caption{Frequency histogram for BD-rate gains of OSLO over 2D learning in ERP. Top row represents SPSNR and WSPSNR BD-rate gains for the architecture of \cite{balle_end--end_2017}. Bottom row represents the results for \cite{balle_variational_2018}.}
\label{fig:BD_rate_hist} 
\end{figure}

In addition to these aggregated plots, the Bjontegaard-Delta (BD) rate \cite{bjontegaard2001calculation} is also calculated separately for each test image, allowing us to measure bitrate reduction while maintaining the same quality. More precisely, for each test image RD curve the BD-rate between 2D learning method on ERP (as the reference) and our proposed OSLO is calculated. Negative values of the BD rate represent percentage saving over the reference. For the architecture of \cite{balle_end--end_2017}, OSLO gains -18.3\% and -12.4\% in terms of S-PSNR and WS-PSNR respectively. Regarding the architecture of \cite{balle_variational_2018}, the average BD-rate is -13.7\% in terms of S-PSNR and -8.6\% in terms of WS-PSNR, demonstrating OSLO's superiority over ERP. The frequency histograms of BD-rate gains over the test images under SPSNR and WS-PSNR metrics are shown in \cref{fig:BD_rate_hist} for different architectures. The figure clearly shows that OSLO improves the compression performance significantly.

\subsection{Ablation study: hop+n aggregation}\label{subsec:aggregation_experiment}
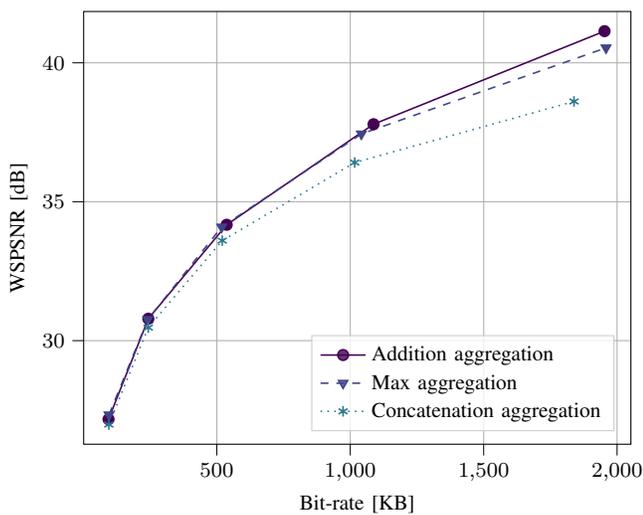
\begin{figure} 
\centering
\fontsize{8pt}{1pt}\selectfont
\setlength\figurewidth{\linewidth}
\setlength\figureheight{0.3\textheight}
\begin{tikzpicture}

\definecolor{color0}{rgb}{0.267004,0.004874,0.329415}
\definecolor{color1}{rgb}{0.253935,0.265254,0.529983}
\definecolor{color2}{rgb}{0.163625,0.471133,0.558148}

\begin{axis}[
height=\figureheight,
legend cell align={left},
legend style={
  fill opacity=0.8,
  draw opacity=1,
  text opacity=1,
  at={(0.97,0.03)},
  anchor=south east,
  draw=white!80!black
},
tick align=outside,
tick pos=left,
width=\figurewidth,
x grid style={white!69.0196078431373!black},
xlabel={Bit-rate [KB]},
xmajorgrids,
xmin=0.79200128173828, xmax=2051.75743109131,
xtick style={color=black},
y grid style={white!69.0196078431373!black},
ylabel={WSPSNR [dB]},
ymajorgrids,
ymin=26.272, ymax=41.848,
ytick style={color=black}
]
\addplot [semithick, color0, mark=*, mark size=2, mark options={solid}]
table {%
94.0177026367188 27.17
243.177618408203 30.79
536.791380615234 34.17
1087.21661621094 37.79
1952.96680175781 41.14
};
\addlegendentry{Addition aggregation }
\addplot [semithick, color1, dashed, mark=triangle*, mark size=2, mark options={solid,rotate=180}]
table {%
95.7849340820313 27.33
240.889793701172 30.76
518.042294921875 34.09
1041.50554077148 37.44
1958.53172973633 40.54
};
\addlegendentry{Max aggregation}
\addplot [semithick, color2, dotted, mark=asterisk, mark size=2, mark options={solid}]
table {%
95.0893176269531 26.98
242.845651855469 30.47
520.150404052734 33.6
1016.58326904297 36.41
1838.59514404297 38.61
};
\addlegendentry{Concatenation aggregation}
\end{axis}

\end{tikzpicture}
\caption{Rate-distortion curves of different aggregation strategies to increase the local support of convolution.}
\label{fig:skip_aggregation} 
\end{figure}

We now evaluate the 3-layer aggregation strategies introduced in Section \eqref{subsec:extension_hop}. To have a fair comparison, we choose the number of feature maps N and M in \cref{fig:spherical_balle} such that all three strategies have almost the same number of learning parameters. Since all experiments use HEALPix sampling, we only consider WS-PSNR as our objective quality metric, which evaluates the performance in the sampling domain. Results are presented in \cref{fig:skip_aggregation}. 
We observe that, for our spherical convolution, the addition aggregation \eqref{eq:sum_skip_connection} outperforms the concatenation \eqref{eq:concat_skip_connection}  and max aggregations \eqref{eq:maxpool_skip_connection}. This justifies the study of aggregation methods, as for graph convolution with regular structure instead, concatenation is shown to be the best aggregation method \cite{xu_representation_2018}.

\subsection{Filter visualization}\label{sec:filter_visualization}
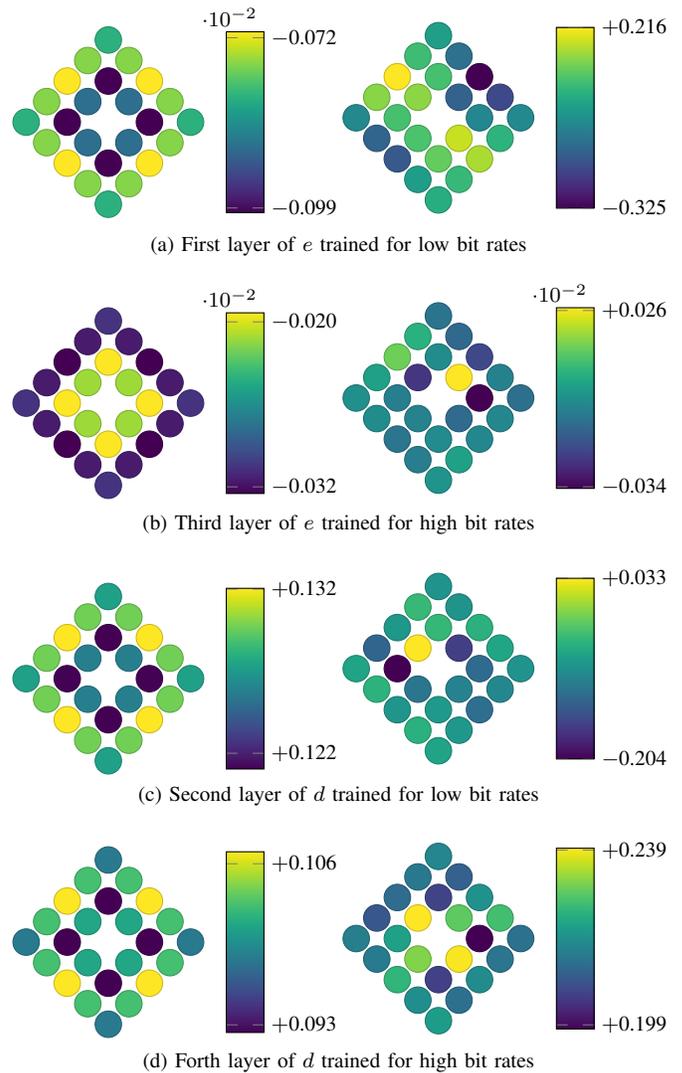
\begin{figure}
\centering
\subfloat[First layer of $e$ trained for low bit rates]{
\fontsize{8pt}{1pt}\selectfont
\setlength\figurewidth{0.45\linewidth}
\setlength\figureheight{0.45\linewidth}
\begin{tikzpicture}

\begin{axis}[
colorbar,
colorbar style={ytick={-0.099,-0.072},yticklabels={
  \ensuremath{-}0.099,
  \ensuremath{-}0.072
},ylabel={}},
colormap/viridis,
height=\figureheight,
hide x axis,
hide y axis,
point meta max=-0.0710548087954521,
point meta min=-0.0997113436460495,
tick align=outside,
tick pos=left,
width=\figurewidth,
x grid style={white!69.0196078431373!black},
xmin=-0.00337476832156435, xmax=0.00337476832156524,
xtick style={color=black},
y grid style={white!69.0196078431373!black},
ymin=-0.00286459304673226, ymax=0.00286459304673226,
ytick style={color=black}
]
\addplot [colormap/viridis, only marks, scatter, scatter src=explicit,mark size=5pt]
table [x=x, y=y, meta=colordata]{%
x  y  colordata
1.59459759391294e-13 0.00260417549702932 -0.08138348162174225
-0.000766990544223135 0.00195312929983574 -0.07621634006500244
0.000766990544462768 0.00195312929974181 -0.07621634006500244
-0.00153398199100865 0.00130208596918541 -0.07105480879545212
7.97296769314758e-14 0.00130208443712445 -0.0997113436460495
0.00153398199116856 0.00130208596899756 -0.07105480879545212
-0.00230097524260338 0.000651043528245246 -0.0762190893292427
-0.000766990544302864 0.000651041996183006 -0.0893920287489891
0.000766990544383038 0.000651041996089077 -0.0893920287489891
0.00230097524268311 0.000651043527963457 -0.0762190893292427
-0.0030679712014221 1.87859055642085e-13 -0.08138623088598251
-0.00153398199108838 9.39293067954001e-14 -0.09970509260892868
0.00153398199108883 -9.39293067954273e-14 -0.09970509260892868
0.00306797120142299 -1.87859055642139e-13 -0.08138623088598251
-0.00230097524268311 -0.000651043527963457 -0.0762190893292427
-0.000766990544382594 -0.000651041996089077 -0.0893920287489891
0.000766990544303308 -0.000651041996183006 -0.0893920287489891
0.00230097524260338 -0.000651043528245246 -0.0762190893292427
-0.00153398199116811 -0.00130208596899756 -0.07105480879545212
-7.97296769314758e-14 -0.00130208443712445 -0.0997113436460495
0.0015339819910091 -0.00130208596918541 -0.07105480879545212
-0.000766990544462324 -0.00195312929974181 -0.07621634006500244
0.000766990544223578 -0.00195312929983574 -0.07621634006500244
-1.59459759391294e-13 -0.00260417549702932 -0.08138348162174225
};
\end{axis}

\end{tikzpicture}

\hspace*{-0.5cm}

\fontsize{8pt}{1pt}\selectfont
\setlength\figurewidth{0.45\linewidth}
\setlength\figureheight{0.45\linewidth}
\begin{tikzpicture}

\begin{axis}[
colorbar,
colorbar style={ytick={-0.325,0.216},yticklabels={
  \ensuremath{-}0.325,
  \ensuremath{+}0.216
},ylabel={}},
colormap/viridis,
height=\figureheight,
hide x axis,
hide y axis,
point meta max=0.216537758708,
point meta min=-0.325507700443268,
tick align=outside,
tick pos=left,
width=\figurewidth,
x grid style={white!69.0196078431373!black},
xmin=-0.00337476832156435, xmax=0.00337476832156524,
xtick style={color=black},
y grid style={white!69.0196078431373!black},
ymin=-0.00286459304673226, ymax=0.00286459304673226,
ytick style={color=black}
]
\addplot [colormap/viridis, only marks, scatter, scatter src=explicit,mark size=5pt]
table [x=x, y=y, meta=colordata]{%
x  y  colordata
1.59459759391294e-13 0.00260417549702932 -0.023134002462029457
-0.000766990544223135 0.00195312929983574 0.047258272767066956
0.000766990544462768 0.00195312929974181 -0.12407515943050385
-0.00153398199100865 0.00130208596918541 0.21653775870800018
7.97296769314758e-14 0.00130208443712445 0.0574357733130455
0.00153398199116856 0.00130208596899756 -0.3255077004432678
-0.00230097524260338 0.000651043528245246 0.12107335031032562
-0.000766990544302864 0.000651041996183006 0.12395710498094559
0.000766990544383038 0.000651041996089077 -0.15588876605033875
0.00230097524268311 0.000651043527963457 -0.20682832598686218
-0.0030679712014221 1.87859055642085e-13 -0.06967344880104065
-0.00153398199108838 9.39293067954001e-14 0.05693084001541138
0.00153398199108883 -9.39293067954273e-14 -0.07789237052202225
0.00306797120142299 -1.87859055642139e-13 -0.070786252617836
-0.00230097524268311 -0.000651043527963457 -0.14965668320655823
-0.000766990544382594 -0.000651041996089077 0.06278948485851288
0.000766990544303308 -0.000651041996183006 0.17235545814037323
0.00230097524260338 -0.000651043528245246 0.025045620277523994
-0.00153398199116811 -0.00130208596899756 -0.1758209764957428
-7.97296769314758e-14 -0.00130208443712445 0.08865493535995483
0.0015339819910091 -0.00130208596918541 0.1470266431570053
-0.000766990544462324 -0.00195312929974181 -0.025519030168652534
0.000766990544223578 -0.00195312929983574 0.036791376769542694
-1.59459759391294e-13 -0.00260417549702932 0.011968460865318775
};
\end{axis}

\end{tikzpicture}
}

\subfloat[Third layer of $e$ trained for high bit rates]{
\fontsize{8pt}{1pt}\selectfont
\setlength\figurewidth{0.45\linewidth}
\setlength\figureheight{0.45\linewidth}
\begin{tikzpicture}

\begin{axis}[
colorbar,
colorbar style={ytick={-0.032,-0.020},yticklabels={
  \ensuremath{-}0.032,
  \ensuremath{-}0.020
},ylabel={}},
colormap/viridis,
height=\figureheight,
hide x axis,
hide y axis,
point meta max=-0.0193397477269173,
point meta min=-0.0324830822646618,
tick align=outside,
tick pos=left,
width=\figurewidth,
x grid style={white!69.0196078431373!black},
xmin=-0.0134997086175224, xmax=0.0134997086175233,
xtick style={color=black},
y grid style={white!69.0196078431373!black},
ymin=-0.0114589550382975, ymax=0.0114589550382975,
ytick style={color=black}
]
\addplot [colormap/viridis, only marks, scatter, scatter src=explicit,mark size=5pt]
table [x=x, y=y, meta=colordata]{%
x  y  colordata
6.37871482431519e-13 0.0104172318529977 -0.030569730326533318
-0.0030679712009437 0.00781277519808927 -0.03152848035097122
0.00306797120190138 0.00781277519771355 -0.03152848035097122
-0.00613600015730379 0.00520850202626698 -0.03248307853937149
3.18922763071126e-13 0.00520840397731255 -0.019339747726917267
0.00613600015794253 0.00520850202551554 -0.03248307853937149
-0.00920414462756949 0.0026042858032849 -0.03152644261717796
-0.00306797120126264 0.00260418775302065 -0.021237481385469437
0.00306797120158245 0.00260418775264493 -0.021237481385469437
0.00920414462788931 0.00260428580215772 -0.03152644261717796
-0.0122724623795658 7.5147158876818e-13 -0.030567696318030357
-0.00613600015762271 3.75721647744313e-13 -0.01935271918773651
0.0061360001576236 -3.75721647744367e-13 -0.01935271918773651
0.0122724623795667 -7.51471588768234e-13 -0.030567696318030357
-0.00920414462788842 -0.00260428580215772 -0.03152644261717796
-0.00306797120158156 -0.00260418775264493 -0.021237481385469437
0.00306797120126353 -0.00260418775302065 -0.021237481385469437
0.00920414462757038 -0.0026042858032849 -0.03152644261717796
-0.00613600015794164 -0.00520850202551554 -0.03248308226466179
-3.18922763071126e-13 -0.00520840397731255 -0.019339747726917267
0.00613600015730467 -0.00520850202626698 -0.03248308226466179
-0.00306797120190049 -0.00781277519771355 -0.03152848035097122
0.00306797120094459 -0.00781277519808927 -0.03152848035097122
-6.37871482431519e-13 -0.0104172318529977 -0.030569730326533318
};
\end{axis}

\end{tikzpicture}

\hspace*{-0.5cm}

\fontsize{8pt}{1pt}\selectfont
\setlength\figurewidth{0.45\linewidth}
\setlength\figureheight{0.45\linewidth}
\begin{tikzpicture}

\begin{axis}[
colorbar,
colorbar style={ytick={-0.034,0.026},yticklabels={
  \ensuremath{-}0.034,
  \ensuremath{+}0.026
},ylabel={}},
colormap/viridis,
height=\figureheight,
hide x axis,
hide y axis,
point meta max=0.026959266513586,
point meta min=-0.0345216318964958,
tick align=outside,
tick pos=left,
width=\figurewidth,
x grid style={white!69.0196078431373!black},
xmin=-0.0134997086175224, xmax=0.0134997086175233,
xtick style={color=black},
y grid style={white!69.0196078431373!black},
ymin=-0.0114589550382975, ymax=0.0114589550382975,
ytick style={color=black}
]
\addplot [colormap/viridis, only marks, scatter, scatter src=explicit,mark size=5pt]
table [x=x, y=y, meta=colordata]{%
x  y  colordata
6.37871482431519e-13 0.0104172318529977 -0.010885068215429783
-0.0030679712009437 0.00781277519808927 0.004428182728588581
0.00306797120190138 0.00781277519771355 -0.013846918009221554
-0.00613600015730379 0.00520850202626698 0.013335734605789185
3.18922763071126e-13 0.00520840397731255 -0.004935970529913902
0.00613600015794253 0.00520850202551554 -0.021340178325772285
-0.00920414462756949 0.0026042858032849 0.0005608024075627327
-0.00306797120126264 0.00260418775302065 -0.02494964376091957
0.00306797120158245 0.00260418775264493 0.026959266513586044
0.00920414462788931 0.00260428580215772 -0.007738269399851561
-0.0122724623795658 7.5147158876818e-13 -0.004111490212380886
-0.00613600015762271 3.75721647744313e-13 -0.008406030014157295
0.0061360001576236 -3.75721647744367e-13 -0.03452163189649582
0.0122724623795667 -7.51471588768234e-13 -0.011972928419709206
-0.00920414462788842 -0.00260428580215772 -0.004886332433670759
-0.00306797120158156 -0.00260418775264493 -0.007266780361533165
0.00306797120126353 -0.00260418775302065 -0.013490969315171242
0.00920414462757038 -0.0026042858032849 -0.006018776912242174
-0.00613600015794164 -0.00520850202551554 -0.010508213192224503
-3.18922763071126e-13 -0.00520840397731255 -0.005407253280282021
0.00613600015730467 -0.00520850202626698 -0.006226133089512587
-0.00306797120190049 -0.00781277519771355 -0.008495209738612175
0.00306797120094459 -0.00781277519808927 0.0004515610635280609
-6.37871482431519e-13 -0.0104172318529977 -0.0027302068192511797
};
\end{axis}

\end{tikzpicture}
}

\subfloat[Second layer of $d$ trained for low bit rates]{
\fontsize{8pt}{1pt}\selectfont
\setlength\figurewidth{0.45\linewidth}
\setlength\figureheight{0.45\linewidth}
\begin{tikzpicture}

\begin{axis}[
colorbar,
colorbar style={ytick={0.122,0.132},yticklabels={
  \ensuremath{+}0.122,
  \ensuremath{+}0.132
},ylabel={}},
colormap/viridis,
height=\figureheight,
hide x axis,
hide y axis,
point meta max=0.132047057151794,
point meta min=0.121040686964989,
tick align=outside,
tick pos=left,
width=\figurewidth,
x grid style={white!69.0196078431373!black},
xmin=-0.0270034843198179, xmax=0.0270034843198184,
xtick style={color=black},
y grid style={white!69.0196078431373!black},
ymin=-0.0229216415210977, ymax=0.0229216415210977,
ytick style={color=black}
]
\addplot [colormap/viridis, only marks, scatter, scatter src=explicit,mark size=5pt]
table [x=x, y=y, meta=colordata]{%
x  y  colordata
1.27595067857056e-12 0.0208378559282707 0.12730082869529724
-0.00613600015666627 0.0156272018759846 0.1296842396259308
0.0061360001585805 0.0156272018752332 0.1296842396259308
-0.0122724623789283 0.0104180163112171 0.13204705715179443
6.37871482431464e-13 0.0104172318529977 0.12104068696498871
0.0122724623802046 0.0104180163097142 0.13204705715179443
-0.0184098488690206 0.00520928652640798 0.12967412173748016
-0.00613600015730423 0.00520850202626698 0.12575717270374298
0.00613600015794253 0.00520850202551554 0.12575717270374298
0.0184098488696586 0.00520928652415342 0.12967412173748016
-0.0245486221089254 1.50316957491581e-12 0.127290740609169
-0.0122724623795662 7.51471588768142e-13 0.12105909734964371
0.0122724623795667 -7.51471588768169e-13 0.12105909734964371
0.0245486221089258 -1.50316957491584e-12 0.127290740609169
-0.0184098488696586 -0.00520928652415342 0.12967412173748016
-0.00613600015794209 -0.00520850202551554 0.12575717270374298
0.00613600015730468 -0.00520850202626698 0.12575717270374298
0.0184098488690206 -0.00520928652640798 0.12967412173748016
-0.0122724623802042 -0.0104180163097142 0.13204705715179443
-6.37871482431464e-13 -0.0104172318529977 0.12104068696498871
0.0122724623789288 -0.0104180163112171 0.13204705715179443
-0.00613600015858005 -0.0156272018752332 0.1296842396259308
0.00613600015666672 -0.0156272018759846 0.1296842396259308
-1.27595067857056e-12 -0.0208378559282707 0.12730082869529724
};
\end{axis}

\end{tikzpicture}

\hspace*{-0.5cm}

\fontsize{8pt}{1pt}\selectfont
\setlength\figurewidth{0.45\linewidth}
\setlength\figureheight{0.45\linewidth}
\begin{tikzpicture}

\begin{axis}[
colorbar,
colorbar style={ytick={-0.204,0.033},yticklabels={
  \ensuremath{-}0.204,
  \ensuremath{+}0.033
},ylabel={}},
colormap/viridis,
height=\figureheight,
hide x axis,
hide y axis,
point meta max=0.0330210477113724,
point meta min=-0.204428970813751,
tick align=outside,
tick pos=left,
width=\figurewidth,
x grid style={white!69.0196078431373!black},
xmin=-0.0270034843198179, xmax=0.0270034843198184,
xtick style={color=black},
y grid style={white!69.0196078431373!black},
ymin=-0.0229216415210977, ymax=0.0229216415210977,
ytick style={color=black}
]
\addplot [colormap/viridis, only marks, scatter, scatter src=explicit,mark size=5pt]
table [x=x, y=y, meta=colordata]{%
x  y  colordata
1.27595067857056e-12 0.0208378559282707 -0.08224523067474365
-0.00613600015666627 0.0156272018759846 -0.044823747128248215
0.0061360001585805 0.0156272018752332 -0.08178591728210449
-0.0122724623789283 0.0104180163112171 -0.0774366706609726
6.37871482431464e-13 0.0104172318529977 -0.0464819073677063
0.0122724623802046 0.0104180163097142 -0.05366634204983711
-0.0184098488690206 0.00520928652640798 -0.12807783484458923
-0.00613600015730423 0.00520850202626698 0.033021047711372375
0.00613600015794253 0.00520850202551554 -0.16067546606063843
0.0184098488696586 0.00520928652415342 -0.06419237703084946
-0.0245486221089254 1.50316957491581e-12 -0.06595665961503983
-0.0122724623795662 7.51471588768142e-13 -0.20442897081375122
0.0122724623795667 -7.51471588768169e-13 -0.12180276215076447
0.0245486221089258 -1.50316957491584e-12 -0.08240772783756256
-0.0184098488696586 -0.00520928652415342 -0.0511675663292408
-0.00613600015794209 -0.00520850202551554 -0.11202441155910492
0.00613600015730468 -0.00520850202626698 -0.09988327324390411
0.0184098488690206 -0.00520928652640798 -0.09514124691486359
-0.0122724623802042 -0.0104180163097142 -0.08083179593086243
-6.37871482431464e-13 -0.0104172318529977 -0.07640721648931503
0.0122724623789288 -0.0104180163112171 -0.11959107965230942
-0.00613600015858005 -0.0156272018752332 -0.05397147312760353
0.00613600015666672 -0.0156272018759846 -0.07816621661186218
-1.27595067857056e-12 -0.0208378559282707 -0.06735487282276154
};
\end{axis}

\end{tikzpicture}
}

\subfloat[Forth layer of $d$ trained for high bit rates]{
\fontsize{8pt}{1pt}\selectfont
\setlength\figurewidth{0.45\linewidth}
\setlength\figureheight{0.45\linewidth}
\begin{tikzpicture}

\begin{axis}[
colorbar,
colorbar style={ytick={0.093,0.106},yticklabels={
  \ensuremath{+}0.093,
  \ensuremath{+}0.106
},ylabel={}},
colormap/viridis,
height=\figureheight,
hide x axis,
hide y axis,
point meta max=0.106967896223068,
point meta min=0.0923838391900063,
tick align=outside,
tick pos=left,
width=\figurewidth,
x grid style={white!69.0196078431373!black},
xmin=-0.00674960017338596, xmax=0.00674960017338596,
xtick style={color=black},
y grid style={white!69.0196078431373!black},
ymin=-0.0057292443750438, ymax=0.0057292443750438,
ytick style={color=black}
]
\addplot [colormap/viridis, only marks, scatter, scatter src=explicit,mark size=5pt]
table [x=x, y=y, meta=colordata]{%
x  y  colordata
3.18922763071237e-13 0.00520840397731255 0.09832156449556351
-0.00153398199084919 0.00390628439868964 0.10264942049980164
0.00153398199132757 0.00390628439850178 0.10264942049980164
-0.00306797120126308 0.00260418775302065 0.10696789622306824
1.5945997623149e-13 0.00260417549702932 0.09238383919000626
0.00306797120158245 0.00260418775264493 0.10696789622306824
-0.00460197485030309 0.00130209822521767 0.10264481604099274
-0.00153398199100865 0.00130208596918541 0.1010197103023529
0.00153398199116811 0.00130208596899756 0.1010197103023529
0.004601974850463 0.00130209822465409 0.10264481604099274
-0.0061360001576236 3.75721647744497e-13 0.09831696003675461
-0.00306797120142254 1.87859055642132e-13 0.0923893079161644
0.00306797120142299 -1.87859055642159e-13 0.0923893079161644
0.0061360001576236 -3.75721647744497e-13 0.09831696003675461
-0.00460197485046255 -0.00130209822465409 0.10264481604099274
-0.00153398199116811 -0.00130208596899756 0.1010197103023529
0.00153398199100865 -0.00130208596918541 0.1010197103023529
0.00460197485030354 -0.00130209822521767 0.10264481604099274
-0.003067971201582 -0.00260418775264493 0.10696789622306824
-1.59459542551131e-13 -0.00260417549702932 0.09238383919000626
0.00306797120126353 -0.00260418775302065 0.10696789622306824
-0.00153398199132757 -0.00390628439850178 0.10264942049980164
0.00153398199084919 -0.00390628439868964 0.10264942049980164
-3.18922763071237e-13 -0.00520840397731255 0.09832156449556351
};
\end{axis}

\end{tikzpicture}

\hspace*{-0.5cm}

\fontsize{8pt}{1pt}\selectfont
\setlength\figurewidth{0.45\linewidth}
\setlength\figureheight{0.45\linewidth}
\begin{tikzpicture}

\begin{axis}[
colorbar,
colorbar style={ytick={0.199,0.239},yticklabels={
  \ensuremath{+}0.199,
  \ensuremath{+}0.239
},ylabel={}},
colormap/viridis,
height=\figureheight,
hide x axis,
hide y axis,
point meta max=0.239433214068413,
point meta min=0.198069885373116,
tick align=outside,
tick pos=left,
width=\figurewidth,
x grid style={white!69.0196078431373!black},
xmin=-0.00674960017338596, xmax=0.00674960017338596,
xtick style={color=black},
y grid style={white!69.0196078431373!black},
ymin=-0.0057292443750438, ymax=0.0057292443750438,
ytick style={color=black}
]
\addplot [colormap/viridis, only marks, scatter, scatter src=explicit,mark size=5pt]
table [x=x, y=y, meta=colordata]{%
x  y  colordata
3.18922763071237e-13 0.00520840397731255 0.21702639758586884
-0.00153398199084919 0.00390628439868964 0.21461041271686554
0.00153398199132757 0.00390628439850178 0.2107483595609665
-0.00306797120126308 0.00260418775302065 0.21258918941020966
1.5945997623149e-13 0.00260417549702932 0.2064596712589264
0.00306797120158245 0.00260418775264493 0.21883071959018707
-0.00460197485030309 0.00130209822521767 0.20932747423648834
-0.00153398199100865 0.00130208596918541 0.23943321406841278
0.00153398199116811 0.00130208596899756 0.22920705378055573
0.004601974850463 0.00130209822465409 0.22700850665569305
-0.0061360001576236 3.75721647744497e-13 0.21577386558055878
-0.00306797120142254 1.87859055642132e-13 0.22156408429145813
0.00306797120142299 -1.87859055642159e-13 0.19806988537311554
0.0061360001576236 -3.75721647744497e-13 0.2134900540113449
-0.00460197485046255 -0.00130209822465409 0.21516749262809753
-0.00153398199116811 -0.00130208596899756 0.2316776067018509
0.00153398199100865 -0.00130208596918541 0.23907846212387085
0.00460197485030354 -0.00130209822521767 0.21412967145442963
-0.003067971201582 -0.00260418775264493 0.22515660524368286
-1.59459542551131e-13 -0.00260417549702932 0.20597180724143982
0.00306797120126353 -0.00260418775302065 0.21806533634662628
-0.00153398199132757 -0.00390628439850178 0.21813608705997467
0.00153398199084919 -0.00390628439868964 0.21331150829792023
-3.18922763071237e-13 -0.00520840397731255 0.2207818329334259
};
\end{axis}

\end{tikzpicture}
}
\caption{Random selection of learned filters at different layers of $e$ and $s$. The weight corresponding to the central pixels is not shown in the filter for better visibility. Left column represents filters learned with DeepSphere that results in isotropic filters. Right column represents the filters learned with OSLO solution. Two filters on the same row are taken from the exact same position in the architecture.}
\label{fig:filter_visualization} 
\end{figure}

\begin{figure*} 
\centering
\subfloat[2408.43 KB]{
		\centering
        \includegraphics[width=0.45\linewidth]{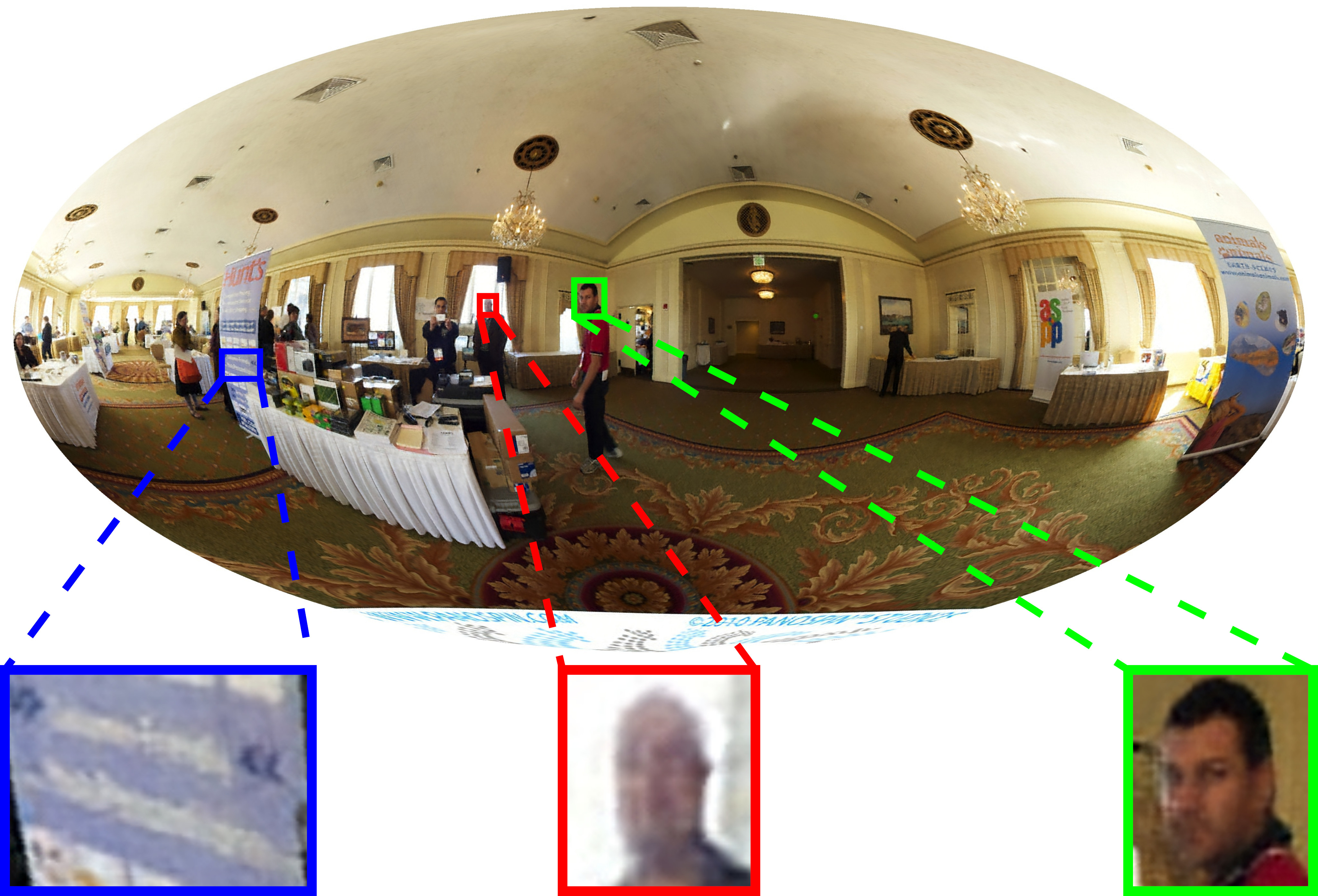}}
\hfill
\subfloat[936.65 KB]{
		\centering
        \includegraphics[width=0.45\linewidth]{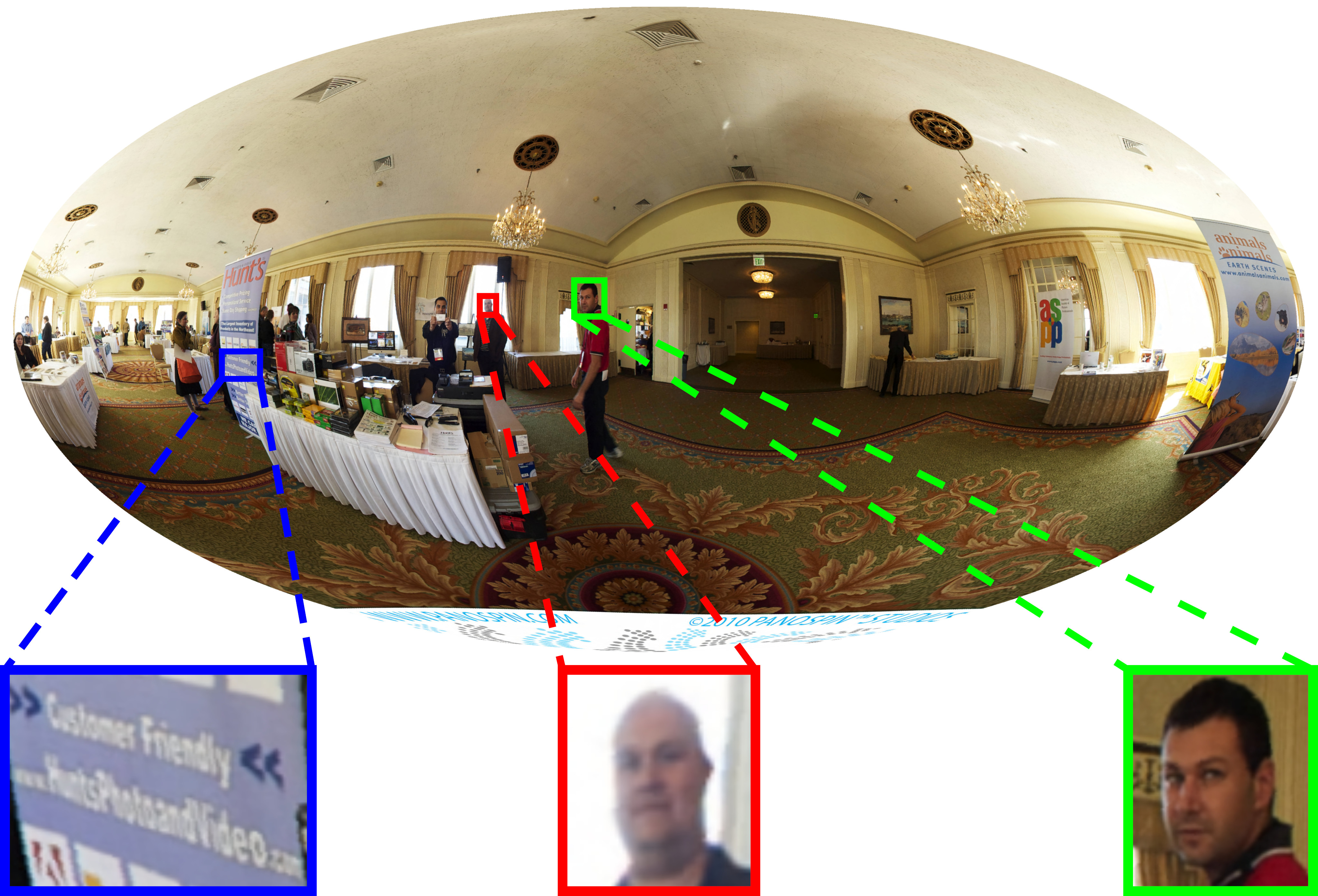}}
\caption{Decompressed images in Mollweide projection. The zoomed versions of red, green, and blue rectangular regions are shown in the bottom row. (a) DeepSphere result stored in 2408.43 KB with 29.54 dB and 29.51 dB for SPSNR and WSPSNR respectively. (b) OSLO result stored in 936.65 KB with 40.91 dB and 41.65 dB for SPSNR and WSPSNR respectively.}
\label{fig:deepsphere_vs_oslo_reconstruction} 
\end{figure*}

In this experiment, we compare our learned filters to those generated from the DeepSphere solution (using graph convolution). We recall that both use the same HEALPix sampling. We randomly select filters at different layers of the network and visualize them in \cref{fig:filter_visualization}. Graph convolution results isotropic filters where each pixel at the same distance from the central one is filtered with the same weight. Such a filter shape obviously limits the expressiveness of the feature maps, which is especially important in image compression. On the contrary, OSLO provides anisotropic filters and can thus be more expressive. To evaluate the impact of this filter's expressiveness on the reconstructed image, we show in \cref{fig:deepsphere_vs_oslo_reconstruction}, the decoded image with both DeepSphere and OSLO-based solutions. It is clear that our solution preserves the details of the image and requires a much lower number of bits.

%
%

\section{conclusion}

We proposed a new representation learning framework for omnidirectional images. The efficient learning operators of 2D CNN models for regular images are redefined on-the-sphere, exploiting the benefits of HEALPix sampling. The proposed OSLO solution allows reaching the expressiveness of a 2D filter while having the advantage of uniform sampling on the sphere, which preserves a low level of complexity. We have finally applied the OSLO solution to the omnidirectional image compression problem by adapting state-of-the-art end-to-end image compression architectures to work on the sphere. Our experiments have shown that our on-the-sphere strategy leads to impressive coding gains when compared with mapping-based or graph-based solutions and demonstrate the promising potential of our OSLO framework for the representation of spherical data, which goes beyond the compression application. Future work includes instantiating OSLO in other contexts for deep learning-based processing of omnidirectional images.



\ifCLASSOPTIONcaptionsoff
  \newpage
\fi



\bibliographystyle{IEEEtran}
\bibliography{main}

\end{document}